\documentclass[apjl,twocolumn]{emulateapj}
\usepackage{epsfig,apjfonts,mathptmx}

\usepackage{natbib}
\usepackage[colorlinks,citecolor=blue]{hyperref} 
\usepackage{url}
\usepackage{graphicx}
\usepackage{float}
\usepackage{placeins} 
\def\gtsima{$\; \buildrel > \over \sim \;$}
\def\ltsima{$\; \buildrel < \over \sim \;$}
\def\prosima{$\; \buildrel \propto \over \sim \;$}
\def\gsim{\lower.5ex\hbox{\gtsima}}
\def\lsim{\lower.5ex\hbox{\ltsima}}
\def\simgt{\lower.5ex\hbox{\gtsima}}
\def\simlt{\lower.5ex\hbox{\ltsima}}
\def\simpr{\lower.5ex\hbox{\prosima}}

\def\eeq{\end{equation}}
\def\beq{\begin{equation}}
\def\24mu{24\,$\mu{\rm m}$}
\def\70mu{70\,$\mu{\rm m}$}
\def\8mu{8\,$\mu{\rm m}$}

\shorttitle{Cold dust in star-forming galaxies at $z=3.62-5.85$}
\shortauthors{Jin et al.}

\begin{document}

\title{Discovery of four apparently cold dusty galaxies at $\lowercase{z}=3.62$--5.85 in the COSMOS field:\\ direct evidence of CMB impact on high-redshift galaxy observables}

\author{
S. Jin\altaffilmark{1,2,3,4},
E. Daddi\altaffilmark{2},
G. E. Magdis\altaffilmark{5,6,7},
D. Liu\altaffilmark{8},
E. Schinnerer\altaffilmark{8},
P. P. Papadopoulos\altaffilmark{9,10},
Q. Gu\altaffilmark{1},
Y. Gao\altaffilmark{11}
and A. Calabr\`o\altaffilmark{2}
}

\altaffiltext{1}{School of Astronomy and Space Science, Nanjing University, Nanjing 210093, China; \textcolor{blue}{shuowen.jin@gmail.com}}
\altaffiltext{2}{CEA, IRFU, DAp, AIM, Universit\'e Paris-Saclay, Universit\'e de Paris,  CNRS, F-91191 Gif-sur-Yvette, France}
\altaffiltext{3}{Instituto de Astrof\'isica de Canarias (IAC), E-38205 La Laguna, Tenerife, Spain}
\altaffiltext{4}{Universidad de La Laguna, Dpto. Astrof\'isica, E-38206 La Laguna, Tenerife, Spain}
\altaffiltext{5}{Cosmic Dawn Center at the Niels Bohr Institute, University of Copenhagen and DTU-Space, Technical University of Denmark}
\altaffiltext{6}{Niels Bohr Institute, University of Copenhagen, DK-2100 Copenhagen, Denmark}
\altaffiltext{7}{Institute for Astronomy, Astrophysics, Space Applications and Remote Sensing, National Observatory of Athens, 15236, Athens, Greece}
\altaffiltext{8}{Max Planck Institute for Astronomy, Konigstuhl 17, D-69117 Heidelberg, Germany}
\altaffiltext{9}{Department of Physics, Section of Astrophysics, Astronomy and Mechanics, Aristotle University of Thessaloniki, Thessaloniki, Macedonia, 54124, Greece}
\altaffiltext{10}{Research Center for Astronomy, Academy of Athens, Soranou Efesiou 4, GR-115 27 Athens, Greece}
\altaffiltext{11}{Purple Mountain Observatory/Key Laboratory for Radio Astronomy, Chinese Academy of Sciences, 8 Yuanhua Road, Nanjing 210034, China}

\begin{abstract}

We report Atacama Large Millimetre Array (ALMA) observations of four high-redshift dusty star-forming galaxy candidates selected from far-Infrared (FIR)/submm observations in the COSMOS field. We securely detect all galaxies in the continuum and spectroscopically confirm them at z=3.62--5.85 using ALMA 3mm line scans, detecting multiple CO and/or [CI] transitions. This includes the most distant dusty galaxy currently known in the COSMOS field, ID85001929 at z=5.847. These redshifts are lower than we had expected as these galaxies have substantially colder dust temperatures (i.e., their SEDs peak at longer rest frame wavelengths) than most literature sources at z>4. 
The observed cold dust temperatures are best understood as evidence for optically thick dust continuum in the FIR, rather than the result of low star formation efficiency with rapid metal enrichment.
We provide direct evidence that, given their cold spectral energy distributions, CMB plays a significant role biasing their observed Rayleigh-Jeans (RJ) slopes to unlikely steep values and, possibly, reducing their CO fluxes by a factor of two. We recover standard RJ slopes when the CMB contribution is taken into account. High resolution ALMA imaging shows compact morphology and evidence for mergers. This work reveals a population of cold dusty star-forming galaxies that were under-represented in current surveys, and are even colder than typical Main Sequence galaxies at the same redshift. 
High FIR dust optical depth might be a widespread feature of compact starbursts at any redshift.
\end{abstract}


\section{Introduction}

Dust temperature is an important parameter of the interstellar medium (ISM) of galaxies, encoded in the shape of their infrared spectral energy distribution (SED) \citep{Draine2007SED}. It is strongly correlated with the mean intensity of the radiation field $\left<U\right>$, that in turn can be used to infer the metallicity-weighted star formation efficiency (i.e., SFE/Z, \citealt{Magdis2012SED}).

It is now well established that for main-sequence (MS) galaxies\footnote{Here and in the rest of the paper we are discussing massive galaxies, above $10^{10}M_\odot$.}, which dominate cosmic star formation at all observed redshifts \citep{Rodighiero2011,Schreiber2015}, the dust temperature is rising with redshift up to at least $z\sim4$.
From the analysis of infrared SEDs at mid-infrared to millimeter wavelengths for individual MS galaxies and for stacked ensembles, 
\cite{Magdis2012SED} were first to measure the evolution of dust temperature with redshift, finding that MS galaxies have more intense radiation fields and thus warmer temperatures as the redshift increases from $z=0$ to 2. 
By stacking far-infrared and submillimetre data from the Herschel Space Observatory, \cite{Magnelli2014} studied the evolution of the dust temperature of MS galaxies up to $z\sim2$, largely confirming these results.
Later on, \cite{Bethermin2015}  found that the mean intensity of the radiation field $\left<U\right>$ increases with increasing redshift up to $z=4$ in MS galaxies, following the trend $(1+z)^{1.8}$ (somewhat faster than in \citealt{Magdis2012SED}) and consistent with models that account for the decrease in the gas metallicity with redshift.
Using both individual detections and stacks of Herschel and Atacama Large Millimetre Array (ALMA) imaging, \cite{Schreiber2018Tdust} also found a consistent trend of increasing dust temperature with redshifts up to $z\sim4$ for MS galaxies. The typical scatter across the MS at fixed redshift is about 0.2-0.3~dex in terms of the intensity of the radiation field $\left<U\right>$ and 10-13\% in terms of $T_{\rm dust}$ \citep{Magdis2012SED,Schreiber2018Tdust}.
{Similar trends of dust temperature have been reported by simulations \citep{Liang_2019Tdust}.}

On the other hand,  starburst galaxies (SBs) have a  different trend of dust temperature with redshift with respect to MS galaxies, and show an almost constant $\left<U\right>\sim 30$ from $z=0$ to $z\sim2$ and perhaps further \citep{Bethermin2015, Schreiber2018Tdust}, when fitted with optically thin dust models like MS galaxies. These SBs are galaxies situated substantially above the MS and are typically (U)LIRGs in the local Universe. At moderate redshifts (up to $z\sim1$--2), SB galaxies are observed to have hotter dust and higher star formation efficiency (SFE) with respect to MS galaxies.
Ensemble average dust temperatures for SBs are not currently  measured with good accuracy at $z>2$. However, 
quite intriguingly, the extrapolation of the trends found by \cite{Bethermin2015} suggests that dust in typical SBs might become colder than MS galaxies starting somewhere at $z>2.5$, and possibly all the way to higher redshifts if the dust temperature-redshift trends discussed above persist in the earlier Universe.  It is currently unclear what could cause such a reversal from the local Universe, where ULIRGs are much warmer than spirals and display much higher star formation efficiencies (SFEs), as a result of the merger process (e.g., \citealt{Renaud2018,Renaud2019arXiv}).

A case of a fairly cold starburst at high redshift has already been observed, individually, in the GN20 galaxy ($z=4.05$, \citealt{Daddi2009GN20,Tan2014}). Its dust temperature has been estimated as $T_{\rm dust}=33$~K \citep{Magdis2012SED}, colder than the average dust temperature of MS galaxies at the same redshift. 
Still, most super-luminous dusty systems at $z>4$ are typically found to show fairly hot dust temperature \citep[$T_{\rm dust}\sim 40-70$~K, e.g.,][]{Riechers2013Nature,Smolcic2015,Riechers2017,Pavesi2018}.   
While dusty galaxies colder than GN20 have not been seen yet at $z>4$,
it is unclear whether the GN20 case is just a curiosity, or the tip of the iceberg for a cold galaxy population missing in current studies.
Thus, investigating galaxy SEDs on larger samples at $z>4$ is fundamental to answer this question.
Although hundreds of square degrees have been mapped at FIR/(sub)mm wavelengths to sufficient depths that should allow detection of dusty star-forming galaxies (DSFGs) with a roughly fixed SFR threshold up to $z \sim5$--10 (particularly from Herschel SPIRE and submm/mm ground based detectors, \citealt{Zavala2017}), 
only a handful of sources have been spectroscopically confirmed to lie at $z > 5$ \citep{Capak2011Nature,Walter2012,Vieira2013,Smolcic2015,Riechers2017,Pavesi2018} and only three at $z > 6$ \citep{Riechers2013Nature,Strandet2017,Zavala2017,Fudamoto2017}.
These samples were pre-selected by their red colors in far-infrared bands \citep{Riechers2013Nature,Riechers2017} and/or strong detections at (sub)mm wavelengths \citep[e.g.,][]{Walter2012,Strandet2017}. 
The SED shapes and FIR colors of galaxies  are subject to the well known degeneracy between dust temperature and redshift, so that the same far-IR SED can be reproduced either by relatively hot dust continuum emission at high redshift, or by intrinsically colder dust  at lower redshift. 
Nevertheless, the spectroscopically confirmed samples at $z>4$--5 are all displaying fairly hot dust content, and cold dusty samples similar to GN20 have not been reported yet.

The persistent sparsity of these very high-z SMG samples is likely not only due to the intrinsic rarity of massive dusty galaxies in the  early Universe, going along with the decrease in the star formation rate density to early times \citep{Lilly1996,Madau1998,Madau2014a,Liu_DZ2017}, but also to incompleteness, i.e., missing detections of more typical objects at faint fluxes in heavily blended FIR/(sub)mm images. 
To counter these problems a number of new generation FIR catalogs have been built \citep{Bethermin_2010_FASTPHOT,Bethermin2012,Bethermin2015,Roseboom2010,Elbaz2011,Safarzadeh2015,Hurley2016,Lee2013}, that should allow in principle selection of DSFGs down to lower luminosities and up to highest redshifts. Notably, we have been developing new 'super-deblended'  catalogs \citep{Liu_DZ2017, Jin2018cosmos} that provide state-of-the-art FIR photometry with well behaved quasi-Gaussian uncertainties while limiting as much as possible the effects of blending from the poor IR PSFs of current ground-based facilities. Our super-deblended catalogs detect a substantial number of $z>4$ FIR-detected galaxy candidates for which only a small fraction of spectroscopic identifications are as yet available. 
Given the less extreme luminosities of these candidates, it is more difficult to identify their redshifts spectroscopically. 
Detection of rest-frame UV/optical lines is time-consuming \citep{Chapman2005_SMG,Capak2011Nature}, and infeasible for DSFGs that are invisible in HST imaging \citep[e.g.,][]{Walter2012,Riechers2013Nature}.
IR/(sub)mm fluxes and FIR lines are easier accessible in these objects than optical features (partially by construction), so that (sub)mm line scans (e.g., CO, [CII]) are more efficient for confirming redshifts for IR-selected star-forming samples at $z>$5--7 \citep{Walter2012,Riechers2013Nature,Vieira2013,Strandet2017,Fudamoto2017}. This method is becoming more prevalent in the ALMA era although the lower luminosity galaxies still require comparably longer efforts for succeeding. 

For cold sources in particular, there are extra difficulties stemming not only from the intrinsically fainter luminosities, but also from the effects caused by the cosmic microwave background (CMB) at high redshift.
The CMB, increasing in temperature proportionally to $(1+z)$, provides an additional source of heating to the ISM, thus increasing the CO luminosities for high-rotational number transitions (e.g., \citealt{Silk1997}). However, taking into account the higher background against which the respective lines must be detected, it actually suppresses any such boost \citep{Papadopoulos2000}.
By modeling CO line emission in starburst galaxies,
\cite{Combes1999} found that the hotter CMB at high-z does not help the detection of CO lines and the net effect is  negative.
For Milky-Way (MW) type galaxies, \cite{Obreschkow2009} found that the CO lines will be dramatically suppressed by the weak contrast against the CMB at $z>5$ and might become totally undetectable at $z\sim7$. 
\cite{daCunha2013} quantified the CMB effects on (sub-)millimeter observations of high-redshift galaxies at $z>5$, and concluded that the inferred dust and molecular gas masses can be severely underestimated for cold systems if the impact of the CMB is not properly taken into account. 
\cite{Zhang_ZY2016} studied the CMB effect on the observed structural and dynamical characteristics of galaxies. They found that elevated CMB at high-z { can dramatically shrink the emergent continuum and low-J CO brightness distributions} of the cold molecular gas, thus erasing spatial and spectral contrasts between their brightness distributions and the CMB. This can strongly affect the measurements of dust and ${\rm H_2}$ gas distribution scale-lengths and velocity fields, and thus the enclosed dynamical masses in galaxies.

It is clear  that the potentially dramatic effects of the CMB, and the resulting biases against the coldest sources, will need to be considered carefully when designing and conducting observations. 
At the same time,  studying cold dusty galaxies at higher-z, discovering more examples of them, will be providing the chance to obtain observational confirmation of the predicted CMB effects, while at the same time helping to clarify the cosmic evolution of dust temperature in galaxies with less biases. This is crucial for constraining the physics of star formation and galaxy formation in the very early Universe.

In this paper, we report the discovery of four remarkably cold dusty galaxies at $z=3.62-5.85$. These galaxies were selected in the FIR/(sub)mm super-deblended catalog in the COSMOS field \citep{Jin2018cosmos}. We present here their spectroscopic observations with ALMA Band 3 scans which confirm their redshifts and allow a detailed investigation of their physical properties. 
The paper is organized as follow. We discuss the targets selection and ALMA observations in Sect.2. The analysis of ALMA spectra, line identifications and redshift estimates are discussed in Sect.3. Sect.4 focuses on the derivation of physical parameters for our target galaxies, with an evaluation of the impact of the CMB on observables. We discuss the implication of these results in Sect.5 and provide our summary and conclusions in Sect.6.
We use standard cosmology (73, 0.23, 0.73) and a Chabrier IMF.

{We emphasize here that throughout the paper we discuss the coldness of the SED as an objective, directly measured property, reflected in the wavelength location of the peak of the spectral energy distribution, in the rest frame. This is thus independent of any physical, model dependent estimate of the actual effective dust temperature. In fact, an hypothesis we will address in the discussion of this paper is that many of these cold sources are only {\em seemingly} cold due to optical depth, while their physical dust temperatures could be much higher. Regardless of the true dust temperature, the CMB effects depend only on the shape of the SED. }

\begin{figure*}
\centering
\includegraphics[width=0.98\textwidth]{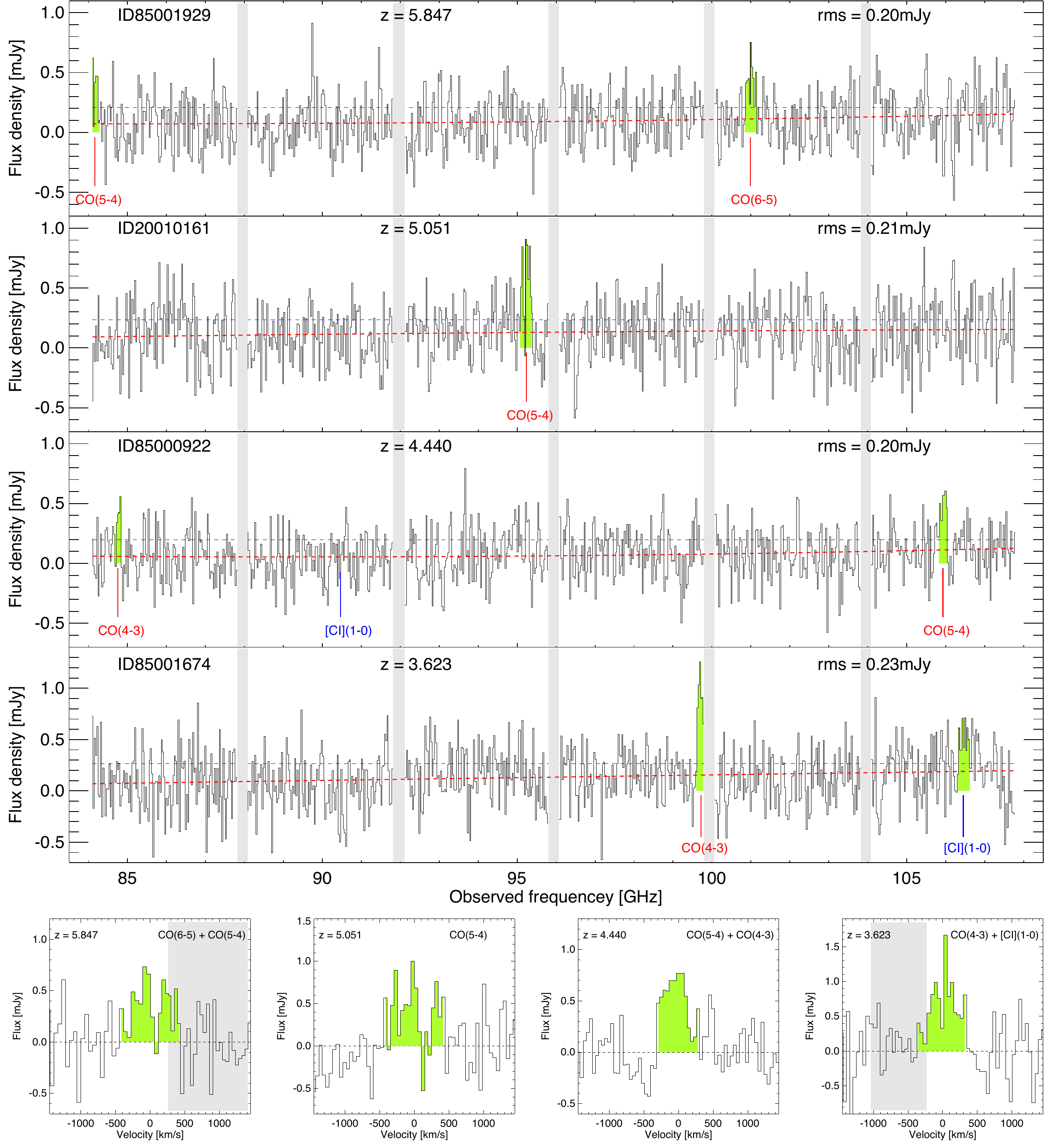}
\caption{
{\it Upper panels}: ALMA 3mm spectra at a resolution of 32~MHz per channel. In order to flatten out noise variation effects (e.g. at sidebands edges), the fluxes are divided by the noise at each channel and multiplied by the average noise of the entire spectrum. Detected lines are highlighted in green and their identifications are labeled.
Grey shaded areas show frequency gaps. 
Blue dashed line indicates the 1$\sigma$ rms noise for each bin, which is also labeled in the upper-right corner, computed for a 32MHz channel, corresponding to 100km~s$^{-1}$ at the average frequency of the spectra.
Red dashed line shows a power-law fit to the continuum going like frequency $\nu^{3.7}$.
{\it Bottom panels}: Stacked line spectra after continuum subtraction. Green areas mark line ranges, the same as in upper panels, as determined by the primary line. Grey areas show the location of gaps in velocity space.
}
\label{spectra}
\end{figure*}

\section{Sample selection, ALMA observations and data reduction}

In the FIR/(sub)mm super-deblended catalog in the COSMOS field, \cite{Jin2018cosmos} selected a sample of 85 $z>4$ high-redshift candidates with {significant IR detection ($\rm S/N_{FIR+mm}>5$ combined)} from  Herschel, SCUBA2, AzTEC and/or MAMBO images \citep{Cowie_2017,Geach2016,Aretxaga2011,Bertoldi2007}.
We singled out four galaxies among these candidates for further ALMA follow-up, based on their highest photometric redshifts $z_{\rm phot, FIR}>6$ that were derived using the $z=6.3$ HFLS3 template \citep{Riechers2013Nature}. 
Among these four galaxies,  ID20010161 was originally detected in the VLA 3~GHz catalog \citep{Smolcic2017}, while the others were found in residual images from the SCUBA2 850~$\mu$m maps from \cite{Geach2016}, after subtracting all known sources in our super-deblending process (see details in Sec. 4.5 in \citealt{Jin2018cosmos}). 
The four sources have no detection in HST nor in UltraVISTA images \citep{McCracken2012UVISTA,Laigle2016}, while we found red IRAC counterparts in the Spitzer Large Area Survey with Hyper-Suprime-Cam  (SPLASH, P. Capak et al., in preparation) imaging datasets,  well-aligned with their VLA/SCUBA2 positions with offsets $<3''$. The subsequent ALMA observations with sub-arcsec beam, described in the next sections, leave no further doubts on the correctness of the IRAC counterpart identification.

We followed-up the four sources with ALMA Band 3 spectroscopic scans in Cycle 5 (Project ID: 2017.1.00373.S, PI: S. Jin).
The Band 3 observations were performed by combining 3 tunings covering 84--108~GHz with resolution of 16~MHz, which is a more  efficient way than performing formal scans albeit at the price of small gaps (850km/s) within the spectral range. The observations were carried out with the array configuration C43-6 giving a synthesized beam of $\sim 0''.85$. The four galaxies were observed with track-sharing, each source was observed for 1.3 hours of on-source time in each tuning, {reaching typical rms sensitivities of $0.1~$mJy per 500~km/s}. We also used ALMA Band 6 imaging (Project ID: 2016.1.00279.S, PI: I. Oteo) available in the archive for two of of our targets. The Band 6 observations are taken in a single tuning at 230~GHz with 56 seconds of integration time per source. One of our sources has also 345~GHz imaging publicly available from the program 2016.1.00463.S (PI: Y. Matsuda). The sources are very strongly detected in the continuum but show no evidence for lines at 230~GHz or 345~GHz.

We processed the 3mm spectra by reproducing the observatory calibration with their custom-made script based on Common Astronomy Software Application package \citep[CASA,][]{McMullin2007CASA}. We converted the data into uvfits format to perform further analysis with the IRAM GILDAS tool working on the uv-space (visibility) data.
The source positions are determined by the centroid of  collapsed 3mm (and 1mm when available) data cubes which mostly show highly significant continuum detections, and used these positions for extracting spectra by fitting  source models in the uv-space. We iteratively searched for evidence of resolved emission either in the continuum, emission line candidates, or combination there-of. Only one source was found to be resolved at more than the 3$\sigma$ level, as discussed below. We further extracted spectra for this source with Gaussian models, and use point-source models for the rest.
The original spectra with noise are extracted with a resolution of 16MHz per channel.
We verified that the noise estimates per channel are reliable from global $\chi^2$ statistics.
We present the binned and noise-normalized spectra in Fig~\ref{spectra}.

We also verified the FIR/(sub)mm photometry of the sources that resulted from SCUBA2 residual maps by obtaining new super-deblended photometry using their better determined positions from ALMA. The difference with the original photometry in \cite{Jin2018cosmos} is entirely negligible.

\begin{table*}
{
\caption{Line detections}
\label{tab:1}
\centering
\begin{tabular}{cccccccccccc}
\hline
     ID       &  $z_{\rm spec}$  &  ${\rm S/N_{1st}}$ & $I_{\rm 1st}$  &   ${N^{\rm EFF}_{\rm 1st}}$ &  $P_{\rm 1st}$ & FWZI &   ${\rm S/N_{2nd}}$ & $I_{\rm 2nd}$ &   ${ N_{\rm 2nd}}$   &   $P_{\rm 2nd}$& $P_{ c, l}$  \\
             &              &               &          [Jy km/s]        &                 &                &  [km/s]            &     &  [Jy km/s]          &              &             &          \\                 
                  \hline
85001929 & 5.847     &    5.22 (CO6-5) & $0.35\pm 0.07$ &  3925 & 7.0e-4  & 885  &    3.27 (CO5-4)  &   $0.22\pm 0.07$ &     12  &     0.013   &   0.9e-5  \\
20010161 & 5.051$^*$    &    5.97 (CO5-4) &  $0.33\pm 0.06$ &   4188 & 1.2e-5 & 850 &    --   &      --   & --      &    -- &  1.2e-5 \\
85000922 &  4.440    &    4.68 (CO5-4) & $0.29\pm 0.06$  &   5162 & 0.014  & 608 &      3.55 (CO4-3)  &  $0.17\pm0.05$  &   13  & 0.005 & 0.7e-4  \\
85001674 &  3.623    &    6.33 (CO4-3) & $0.42\pm 0.07$  &  4294 &  1.3e-6  & 747  &    4.44 ([CI]1-0)  &  $0.25\pm 0.06$  &    13     & 1.4e-4 &   1.8e-10  \\
\hline
\end{tabular}\\
}
{Notes: ${\rm S/N_{1st}}$: significance of the first line; $I_{\rm 1st}$: flux density of the first line; $N^{\rm EFF}_{\rm 1st}$: number of effective trials following Eq.2; $P_{\rm 1st}$: chance probability of the first line; FWZI: full width at zero intensity of the lines; ${\rm S/N_{2nd}}$: significance of the second line; $I_{\rm 2nd}$: flux density of the second line; ${N_{\rm 2nd}}$: number of trials for the second line; $P_{\rm 2nd}$: chance probability of the second line; $P_{ c, l}$: chance probability of 2 lines. $^*$ this redshift is less secure as it is based on single-line identification (see text for details).}
\end{table*}

\begin{figure*}
\centering
\includegraphics[width=0.98\textwidth]{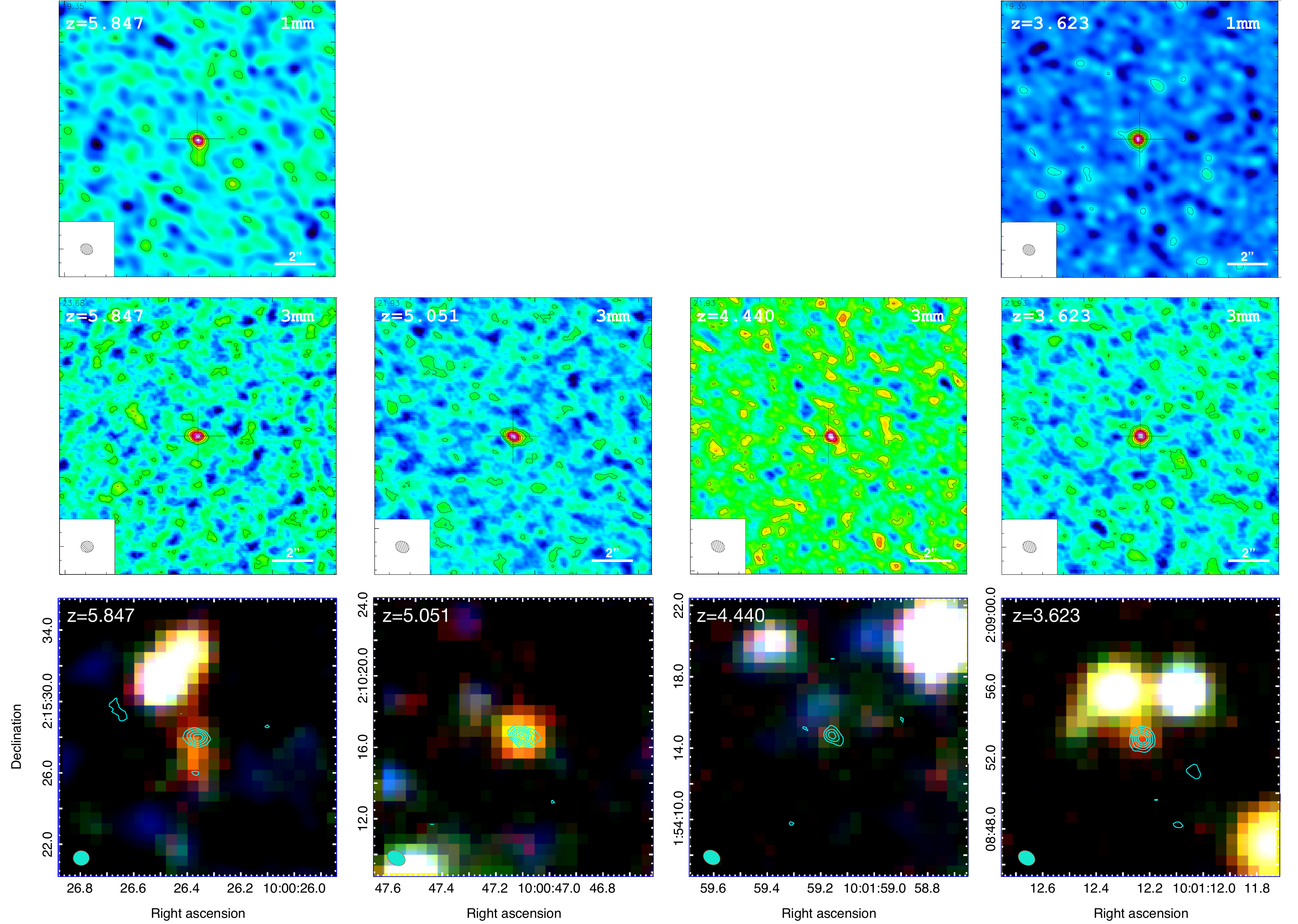}
\caption{%
 ALMA 1~mm  ({\it Upper panels}) and 3mm ({\it Center panels}) images, with contours in step of 2$\sigma$. The size and orientation of the beam is indicated in the bottom left corner.
{\it Bottom panels}: Color images of the 4 galaxies (blue: UltraVISTA $K_s$; green: IRAC $3.6~{\rm \mu m}$; red: IRAC 4.5~${\rm \mu m}$). Cyan contours show ALMA 3~mm continuum starting from 3$\sigma$ in step of 2$\sigma$. 
	}
\label{img}
\end{figure*}

\section{Redshift confirmation}
\label{Sec2} 

By  sweeping through their full 3mm (continuum subtracted) spectrum, we blindly search for the single emission line feature with the highest significance for each galaxy, following the same line-searching algorithm applied in \cite{Daddi2015}, \cite{Coogan2018} and \cite{Puglisi2019arXiv}. This algorithm returns both the SNR  and the optimum velocity range of the line, by determining the pair of starting channel and ending channel, across which the line flux is integrated, that corresponds to the lowest chance probability for the line (hence highest detection significance). The the flux density $I_{\rm 1st}$ (Table~\ref{tab:1}) is integrated over this velocity range which basically corresponds to the FWZI (full width at zero intensity) of the line, which is highlighted in green in Fig.~\ref{spectra} and shown in Table~\ref{tab:1}. Simulations and double-Gaussian fits carried out in \cite{Coogan2018} show that typically this FWZI is very close to the actual FWHM of the line.
As shown in Fig.~\ref{spectra}, the most significant lines in the 4 spectra are detected at 5--6$\sigma$ with full velocity widths of 608--885 km/s (Table~\ref{tab:1}). These linewidths are large but comparable to whose found in the literature for SMGs (e.g., \citealt{Daddi2009GN20,Riechers2013Nature,Riechers2017,Marrone2017Nature}).

\subsection{Chance probability of emission lines}

We run extensive simulations to determine how to compute the chance probability of finding emission lines (with the same line searching algorithm) as produced by noise fluctuations in a given spectral scan,  depending on the SNR of the feature, its linewidth (broader lines are more unlikely to happen by chance because fewer independent realizations are available within the observed range -- see also \citealt{Gonzalez-Lopez2019_3mmSurvey}) and also based on the full velocity range spanned by the data, the velocity sampling, and the range of acceptable linewidths (from 100 to 1000~km~s$^{-1}$ in this case). We  assume well-behaved Gaussian noise in the data, as verified to hold for our ALMA spectra.  
We define the probability that a line detection is due to noise fluctuations as:

\begin{equation}
    P_{\rm line} = 1 - P_{\rm 0}^{N^{\rm EFF}_{\rm trials}}
\end{equation}
where $P_{\rm 0}$ is the probability to have a  line with SNR up to what observed, in a single trial (approaching 1 for high sigma events) and the exponent $N^{\rm EFF}_{\rm trials}$ is the effective number of trials which, based on our simulations, can be well approximated by:

\begin{equation}
    N^{\rm EFF}_{\rm trials} \sim 10 \frac{N_{\rm total,ch}}{N_{\rm line,ch}}N_{\rm line,ch}^{0.58}{\rm log}\frac{N^{\rm max}_{\rm line,ch}}{N^{\rm min}_{\rm line,ch}}
\end{equation}
where $N_{\rm total,ch}$ and $N_{\rm line,ch}$ are the  number of channels in the entire spectrum scanned and in the recovered line, respectively, while $N^{\rm max}_{\rm line,ch}$ and $N^{\rm min}_{\rm line,ch}$ define the allowed velocity range of lines to be searched, expressed as number of channels. Notice that a naive treatment would account only for the first term (ratio of full velocity in spectrum divided by line velocity), while in reality the number of effective realizations is considerably larger. In fact, less than a full line-velocity shift is required to imply a new realization, and  the number of effective realizations depends also on the range of velocity widths searched as the line-width is not known a-priori. We emphasize that this recipe, and in particular Eq.2, holds only in the assumption that there is no active spatial search for the line position, as in our case where the position for the spectral extractions are all strongly constrained by the continuum detections. When one is blindly searching for emission lines having also to determine their spatial position, the number of effective trials would be much larger as Eq.2 would require to be multiplied by the number of effective spatial positions searched \footnote{This is typically assumed to be the area within the primary beam divided by the area covered by half the synthesized beam, although dedicated simulations would be required to confirm this accurately.}.

\subsection{Measuring redshifts for our sample}

We report the spurious probability for the strongest line in each spectrum in Table~1 ($P_{\rm 1st}$). They are below 0.1\% in all cases except for the 4.7$\sigma$ line of ID85000922 that has a 1.4\% probability of being spurious, by itself. 

Then we searched for additional matching lines based on the redshift solutions determined by the first ones, using the same velocity range as determined by the primary line, assuming they are CO transitions.
We excluded, in fact,  solutions in which the primary line was [CI] or H$_{2}$O, given that stronger accompanying CO lines would have been detected in the spectra in those cases.
Meaningful second lines were detected in 3 galaxies with significance of 3.3--4.4$\sigma$. We compute chance probabilities to find matching second lines ($P_{\rm 2nd}$) based on the significance of each line and the number of search trials  performed ($N_{\rm 2nd}$). The combined probability for the two-line match is in all cases $\sim10^{-4}$ or lower, hence we consider these three galaxies as reliable spectroscopic confirmations.
The redshifts (and line identifications) are 
determined as CO(5-4)/CO(6-5) at $z=5.85$, CO(4-3)/CO(5-4) at $z=4.44$ and CO(4-3)/[CI](1-0) at $z=3.62$ (Fig.~\ref{spectra}).
In Fig.~\ref{spectra}-bottom, we present zooms to the weighted averages of  lines detected for each galaxy after continuum subtraction. 
Given that the CO(4-3) line of ID85001674  partially falls in a frequency gap, we ultimately calculate its redshift and velocity width from the [CI](1-0) line. We note that the confirming/second line for ID85001929 is located at the edge of the spectral range, and only partially covered by our data. Still, the available signal within the velocity range pre-determined by the first line is strong enough that its spurious probability is at 0.013. We thus maintain that this  $z=5.85$ identification is secure\footnote{Interestingly, some excess flux at 1.1~mm is found in its FIR SED (Fig.~\ref{sed}), which hints at the presence of a  bright [CII]~158$\mu$m line  at a consistent $z\sim5.9$  redshift that is  boosting the AzTEC and MAMBO photometry. However, to make an impact this would need to be at the level of many dozens Jy*km/s, which would imply $L_{\rm [CII]}$  to $L_{\rm IR}$ ratios at the level of local galaxies $>3\times10^{-3}$. Direct observations are required to confirm this tentative evidence.}.
Therefore, ID85001929 at $z=5.85$  is now the most distant known SMG in the COSMOS field, at a higher redshift than the recently found dusty galaxy CRLE at $z=5.67$ \citep{Pavesi2018}.

The ID20010161 has a single line detection at 95.2~GHz with 5.97$\sigma$ significance and a spurious probability of $\sim10^{-5}$ (Table~\ref{tab:1}), hence it is highly secure. However the redshift identification is less secure, as there are  multiple solutions given the lack of a second line. 
We disfavor the single line identification as  CO(7-6) at $z=7.47$ given the lack of [CI](2-1) (the formal signal within the expected velocity range gives -2.7$\sigma$ at 95.5~GHz) and H$_2$O($2_{11}-2_{02}$) (0.07$\sigma$ at 88.8~GHz) lines, since the H$_2$O line is typically 0.4--1.1 of the high-J CO flux in high-z lensed ULIRGs \citep{Omont2013,Yang_CT2016,Yang_CT2019}. 
We also exclude CO(6-5) at $z=6.26$ because of the lack of detection of H$_2$O($2_{11}-2_{02}$) (-1.075$\sigma$ at 103.6~GHz). 
The solution with CO(4-3) at $z=3.84$ is also less likely due to the lack of any [CI](1-0) signal (-0.7 $\sigma$ at 101.7~GHz), although it is not unconceivable that [CI](1-0) might be intrinsically faint for this source as it can reach $\sim25$\% of the CO(4-3) flux (e.g., \citealt{Walter2011Ci}; Valentino et al 2018; 2019). This redshift is less secure given that no confirming lines are present in its spectrum. 
However, we consider the most likely identification of this line as CO(5-4) at $z=5.05$, for which we would { not} expect other significant lines in the observed range. In the following we will use this redshift identification for this source. If the redshift should turn out to be instead $z=3.84$, its dust temperature would be even lower and the conclusions derived from these sources even further strengthened. 

These redshift identifications consistently confirm that no strong lines are expected in the 230~GHz spectral ranges of the Oteo data nor in the 345~GHz spectral range of the Matsuda data, for either of the two galaxies for which 230~GHz or 345~GHz observations are available.

\section{Results}

\subsection{Multi-wavelength imaging and morphology}

In Fig.~\ref{img}, we present ALMA (clean) images for this sample. 
We solidly detected continuum emission with peak significance of 9--14$\sigma$ at 3~mm for the four sources, { 13--19$\sigma$ at 1mm and 29$\sigma$ at 870~$\mu$m for those in which these are available (Table~\ref{tab:alma}). }
We measured their sizes by fitting models in uv space in the ALMA 1/0.8~mm and 3~mm collapsed datasets (also adding lines when available, but these provide only a modest contribution being at substantially lower SNRs). We combine the different tracers for a single galaxy in the uv space, following the procedure presented in \cite{Puglisi2019arXiv}. Only the lowest redshift galaxy,  ID85001674 at $z=3.62$, is resolved with a FWHM size of $0''.42\pm 0''.04$ ($3.1\pm 0.3$ kpc), implying a SFR surface density ($\Sigma_{\rm SFR}$) of $22_{-9}^{+13}$ ${\rm M_\odot yr^{-1} kpc^2}$. 
The remaining sources are unresolved, we show their $2\sigma$ size upper limits (and lower limits of SFR surface density $\Sigma_{\rm SFR}$) in Table~\ref{tab:2}. 
 ID85000922 at  $z=4.4$ has a size limit of $<2.8$~kpc, and the other two galaxies at $z>5$ appear to be more compact with FWHM sizes $<1.9$~kpc. 
The measured sizes of our ALMA sources are significantly smaller than the 3~GHz sizes reported by \cite{Miettinen2017size} for a sample of 115 known SMGs in the COSMOS field, with a median size of $4.6\pm0.4$~kpc.
In contrast, our measured sizes are close to the ALMA sizes of high-z SMGs from \cite{Ikarashi2015alma} and \cite{Simpson2015}, who report median sizes of $1.6\pm0.14$~kpc at 1.1 mm, and $2.4\pm0.2$~kpc at $870\mu$m, respectively. 
Our ALMA sizes also agree with the 10~GHz-selected sample in \cite{Murphy2017vla10ghz}, which has a median size of $1.20\pm0.28$~kpc.

 \begin{table}[ht!]
{
\caption{ALMA continuum measurements}
\label{tab:alma}
\centering
\begin{tabular}{cccc}
\hline
     ID       &  $S_{\rm 3mm}$ [mJy]&  $S_{\rm 1.3mm}$ [mJy] &   $S_{\rm 870\mu m}$ [mJy] \\
85001929   & $0.088\pm0.007$ & $1.52\pm0.12$ & --\\
20010161  &  $0.115\pm0.008$  & -- &--\\
85000922  & $0.068\pm0.007$  & -- &--\\
85001674  & $0.124\pm0.009$  &   $2.59\pm 0.14$ & $7.48\pm0.26$\\
\hline
\end{tabular}\\
}
\end{table}

The high compactness of this sample is unlikely to be caused by AGN activity, because the emission from any dusty torus would be negligible at FIR/(sub)mm wavelength, and it cannot dominate the 3~mm emission (rest frame $440-650\mu$m for this sample).
In any case, we do not see any evidence for an AGN component in their cold SEDs (see Sec. 5.2) and none is X-ray detected from \citet{Jin2018cosmos}.
Therefore, we speculate that the compact morphology in this very highly star-forming sample is a consequence of ongoing mergers, similarly to local ultra-luminous infrared galaxies (ULIRGs) \citep{Soifer2000,Juneau2009} and high-z compact starbursts \citep{Puglisi2017,Calabro2019,Marrone2017Nature}.  

In Fig.~\ref{img}-bottom, we show $K_{\rm s}$+SPLASH color images for this sample. 
All sources have red counterpart in SPLASH images quite consistent with the ALMA imaging, however, only two of them are entirely consistent in position and extension with the ALMA detections.
The SPLASH counterpart of the $z=5.85$ galaxy appears to be more extended than its ALMA morphology, the IRAC peak being located about $0''.8$ to the south, where some faint extension is seen at 1~mm in the ALMA imaging as well (Fig.~\ref{img}-top-left).
ID85000922 at $z=4.44$  has a very faint SPLASH IRAC 3.6~$\mu$m counterpart at its ALMA position, while the blue source with $\sim 2''.0$ offset appears to be unrelated having a well constrained $z_{\rm phot}=0.86$ \citep{Laigle2016}.

The ID20010161 and ID85001674 galaxies are detected in the 3~GHz image with $5.0\sigma$ and $3.6\sigma$ \citep{Smolcic2017,Daddi2017}, showing infrared-to-1.4~GHz radio luminosity ratio $q_{\rm IR}=2.3$ and 2.4, which agree well with the evolution of $q_{\rm IR}(z) = (2.88\pm 0.03)(1 + z)^{−0.19\pm 0.01}$ for star forming galaxies in \cite{Delhaize2017qIR}, while the other sources are either marginally or not detected in the radio ($<2.8\sigma$).
No signal is observed  in the UltraVISTA $K_s$ band (or any shorter wavelength) for our sources, which thus could be qualified as $K_s$-band dropouts. No photometric redshift is, of course, available for these sources in near-IR or optically selected catalogs.

\begin{figure*}
\centering
\includegraphics[width=0.96\textwidth]{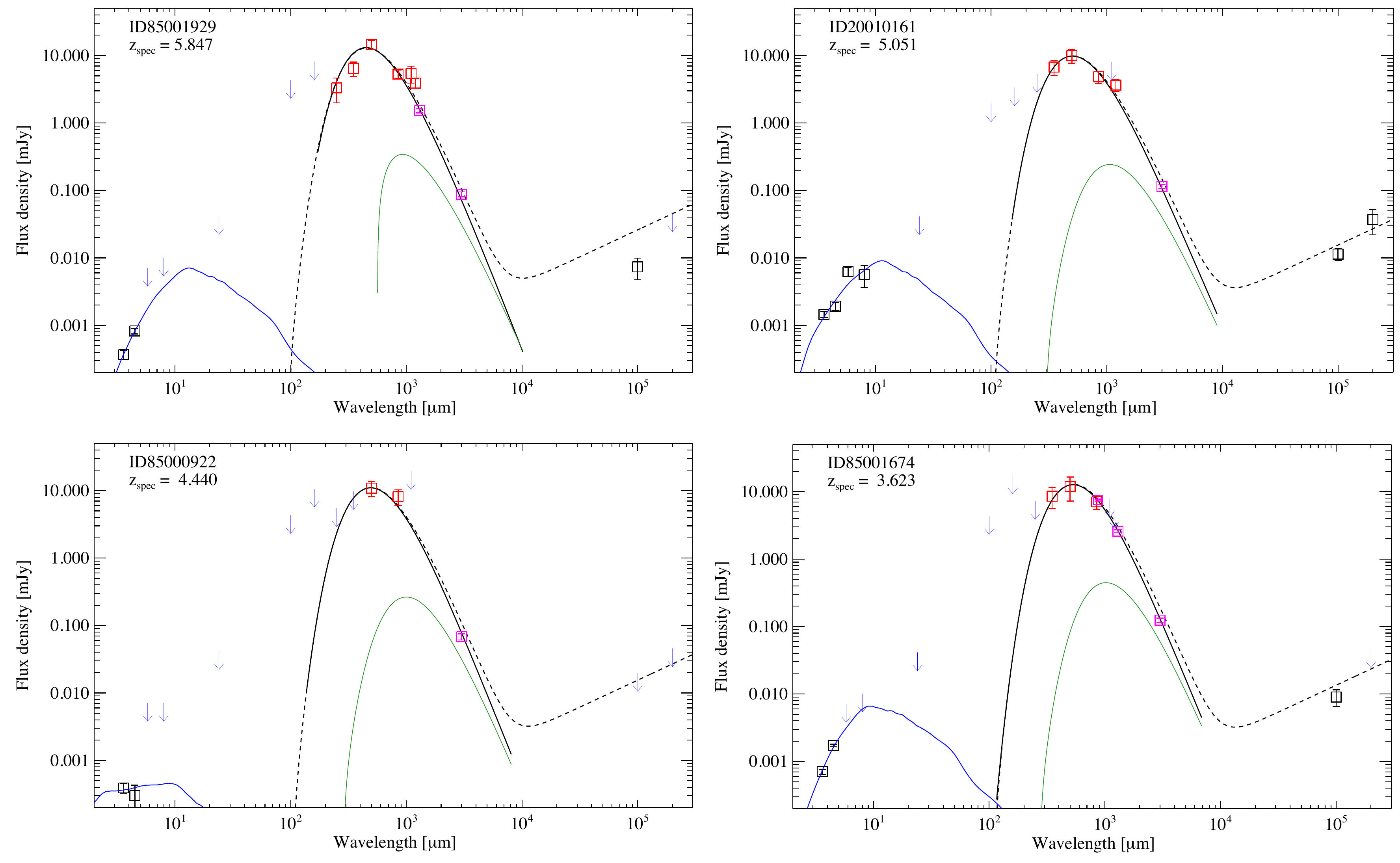}
\caption{
Dust SEDs of the four galaxies in this work.
Photometries are taken from the super-deblended catalog \citep{Jin2018cosmos}, ALMA (magenta) 3~mm, 1~mm and/or $870~\mu$m observations. Arrows mark the 3$\sigma$ upper limits.
The SEDs are fitted by modified black body (MBB) models accounting for CMB impacts on the continuum \citep{daCunha2013}. 
The solid (dashed) curve shows the best fit to the observed SED accounting (not accounting) for CMB. A radio component has been added assuming an evolving $q_{\rm IR}$ from \cite{Delhaize2017qIR}.
The green curves indicate the flux removed from the continuum SEDs by the effects of the CMB. The most impacted fluxes are the ones at 3~mm.
	}
\label{sed}
\end{figure*}

\begin{table*}
{
\caption{CMB impact on observables}
\label{tab:2}
\centering
\begin{tabular}{ccccccccccccc}
\hline\hline
     ID       &   $L_{\rm IR}$  &    $\beta_{\rm thin}^{\rm noCMB}$ &   $\beta_{\rm thin}^{\rm CMB}$ &  $\beta_{\rm thick}^{\rm CMB}$ &   $T_{\rm dust,thin}^{\rm noCMB}$ &   $T_{\rm dust,thin}^{\rm CMB}$ &   $T_{\rm dust,thick}^{\rm CMB}$  & $M_{\rm dust,thin}^{\rm noCMB}$ & $M_{\rm dust,thin}^{\rm CMB}$ & $M_{\rm dust,thick}^{\rm CMB}$ &  $L'^{\rm obs}_{\rm CO(5-4)}$  &   $L'^{\rm CMB}_{\rm CO(5-4)}$  \\
    		  &    [$10^{12}L_\odot$]      &                &             &       &  [K]          &  [K]   & [K]       &  [$10^{8}$M$_\odot$] & [$10^{8}$M$_\odot$]  & [$10^{8}$M$_\odot$] &    (1)    &  (1)  \\
 \hline
85001929    &     12.8$\pm$1.8 &  2.5$\pm$0.2 &    2.2$\pm$0.2   & 2.1$\pm$0.2 &      40$\pm$3     &   42$\pm$3 & 61$\pm$8  &	 3.5$\pm$0.6 &	 4.3$\pm $1.1 & 2.2$\pm$0.9 &  2.0$\pm$0.4$^{(2)}$  & 2.9$\pm$5.5$^{(2)}$ \\
20010161    &      6.2$\pm$1.0  &     2.4$\pm$0.2 &    2.0$\pm$0.2 & 1.9$\pm$0.2   &      32$\pm$3     &   35$\pm$4 &  40$\pm$6  &  8.7$\pm$2.5 & 10.4$\pm$2.9  & 7.7$\pm$2.1 &  1.3$\pm$0.2  & 1.8$\pm$0.3 \\
85000922    &   5.1$\pm$1.0 &     3.7$\pm$0.6  &    2.3   &   2.1$\pm$0.4 &    20$\pm$ 4     &   30$\pm$4 &    42$\pm$6 & 18.3$\pm$9.8	& 10.2$\pm$1.3 & 7.0$\pm$2.3 & 0.9$\pm$0.2  & 1.3$\pm$0.3\\
85001674    &   3.3$\pm$0.8 &    2.7$\pm$0.2 &    2.2$\pm$0.2  & 2.1$\pm$0.2  &      23$\pm$2     &   24$\pm$2 &   41$\pm$5  & 29.2$\pm$8.9	& 33.6$\pm$8.1 & 13.0$\pm$3.7 & 0.8$\pm$0.1$^{(2)}$  & 1.2$\pm$0.2$^{(2)}$ \\
\hline
\hline
\end{tabular}\\
}
{Notes:   
$L_{\rm IR}$: IR luminosity at 8--1000${\rm \mu m}$ corrected for CMB effect (dashed curve in Fig.~\ref{sed}); 
$\beta_{\rm thin}^{\rm noCMB}$, $T_{\rm dust,thin}^{\rm noCMB}$ and $M_{\rm dust,thin}^{\rm noCMB}$: RJ slope, dust temperature and dust mass from optically thin MBB fits without accounting for CMB; 
$\beta_{\rm thin}^{\rm CMB}$, $T_{\rm dust,thin}^{\rm CMB}$ and $M_{\rm dust,thin}^{\rm CMB}$: RJ slope, dust temperature and dust mass from optically thin MBB fits accounting for CMB effect; 
$\beta_{\rm thick}^{\rm CMB}$, $T_{\rm dust,thick}^{\rm CMB}$ and $M_{\rm dust,thick}^{\rm CMB}$: RJ slope, dust temperature and dust mass from optically thick MBB fits accounting for CMB effect; 
$L'^{\rm CMB}_{\rm CO(5-4)}$: CO(5-4) luminosities corrected for CMB effect assuming $T_{\rm exc}=T_{\rm dust,thin}^{\rm CMB}$ in LTE condition.
(1) ${\rm 10^{10}~K~km/s~pc^2}$.
(2) the observed CO(5-4) luminosities for these two sources are extrapolated as $1.2\times L'_{\rm CO(4-3)}$ and $0.8\times L'_{\rm CO(6-5)}$, respectively.
}
\end{table*}

\subsection{Cold dust against the CMB}

We collected super-deblended FIR/(sub)mm photometry together with the ALMA 0.8/1mm and 3mm continuum measurements for deriving their full  dust SEDs. 
We start by fitting these SEDs using modified black-body (MBB) models from \cite{Casey2012} with free $T_{\rm dust}$ and $\beta$, not accounting for the effect of the CMB. We find that these galaxies are fitted with very cold dust temperatures $T_{\rm dust}=20$--41 K and abnormally steep RJ slopes $\beta=2.4$--3.7 which appear to be well determined with relatively small errors, particularly for the two objects with highly accurate 1mm and 3mm ALMA continuum measurements in the RJ tail. Such steep slopes had not been reported in literature previously.
However, galaxies are always observed against the CMB, and the CMB effect on the continuum would be non-negligible on dusty SEDs especially for our systems which appear to be fairly cold for their  high-z \citep{daCunha2013}. 

We thus use alternative {optically thin} MBB SEDs \citep{Magdis2012SED} accounting for the CMB effect on dust continuum following the prescriptions by \cite{daCunha2013}. {MBB models are not ideal to describe the Wien part of galaxies SEDs but are  appropriate here, given the photometric sampling available for our galaxies.}
{We convert the luminosity of MBB models $L_{\rm MBB}$ to  $L_{\rm IR,8-1000\mu m}$  multiplying by a constant of 1.35, a median value for starburst sample \citep{Magdis2012SED} based on comparison with Draine and Li (2007).}
In Fig.~\ref{sed}, we show the best fit to the observed photometry with solid curves, while dashed curves indicate the intrinsic SEDs before accounting for the CMB effect on the continuum (green curve). We can clearly see that the CMB depresses the dust continuum  at longer wavelengths, severely impacting its level at 3~mm {(see Section 4.4)}, thus making the observed RJ slope steeper. 
In Table~\ref{tab:2}, we list derived parameters {for fits with and without accounting for the CMB.} 
In both cases we find consistently cold dust temperature ${T_{\rm dust}=20}$--42 K, which are also consistent with those derived by fitting  \cite{Casey2012} models.
The consistence of ${T_{\rm dust,obs}}$ and ${T_{\rm dust,thin+CMB}}$ shows that accounting for the CMB effect does not alter ${T_{\rm dust}}$ for this sample, implying that the effects of the ${T_{\rm dust}}$-- $\beta$ degeneracy are negligible for our galaxies, within the uncertainties. This is because the CMB is affecting the observed continuum at long enough wavelengths, while it has barely any effect on the peak of the dust SED where ${T_{\rm dust}}$ is measured.
We verified that the impact of the CMB on the derived $T_{\rm dust}$ is negligible out to $z=5$, even for a galaxy with intrinsic $T_{\rm dust}=18$~K, the marginal overestimate of $T_{\rm dust}$ would be less than 1~K in that case.
Strikingly, the SEDs {corrected for CMB effect yield  $\beta\sim$2-2.2 },  consistent with typical $\beta$ values found in the literature for star-forming galaxies, within the uncertainties (of about 0.2 in $\beta$).

Have we thus really directly detected the effect of the CMB on galaxy SEDs, for the first time (as far as we know)? This appears to be the case as 
we argue that the $\beta$ values of order 2.4--3 are not plausible and the CMB is indeed required to obtain a physically meaningful interpretation of the SEDs of our targets. 
In fact, large Herschel surveys have been used to constrain the slopes of local galaxies finding $\beta\approx  1.5$ and up to a maximum of 2 \citep{Boselli2012,Bianchi2013}. None of the local galaxies with an apparent best fitting $\beta>2$ is at more than 1-sigma from $\beta=2$ \citep{Remy-Ruyer2013}.
Also, although our high redshift sample might be somewhat unusual because of particularly cold dust for its redshift (as discussed in more details later on in this paper), there is no evidence for steeper $\beta$ slope in cold galaxies in literature. E.g., our Milky-way ( $T_{\rm dust}=19$~K) is colder than this sample but has no steep beta slope, where the analysis of the full MW dust SED yields $ \beta \approx 1.8 \pm0.2$ \citep{PlanckCol2011XVI}. Without accounting for the CMB, the best constrained galaxies in our sample (not considering ID85000922 given its poor SED\footnote{Note that the galaxy ID85000922 ($z=4.44$) has also an abnormally steep $\beta_{\rm obs}=3.7$ but with very large uncertainty. 
When including the CMB we still  fit its SED with a fixed $\beta=2.3$, which seems required for producing both a good fit to the data and a conservatively cold dust temperature ${T_{\rm dust}=30\pm4}$ K. However, it should be noted that its dusty SED is not well constrained given the weak signal in the Herschel bands and we suspect that the SCUBA2 850~$\mu$m photometry is boosted by noise.}) have $\beta>2$ at significance ranging from 2 to 3.5$\sigma$. Taken together, their average SED is steeper than  $\beta=2$ at about the $5\sigma$ level.
Therefore, we conclude that the steep observed $\beta$ slopes are unambiguous evidence that the CMB is having a measurable effect on our target galaxies, which arises from a severe reduction of the dust continuum observed at 3~mm, or about beyond 500~$\mu$m in the rest-frame.

\begin{figure*}[ht]
\centering
\includegraphics[width=0.96\textwidth]{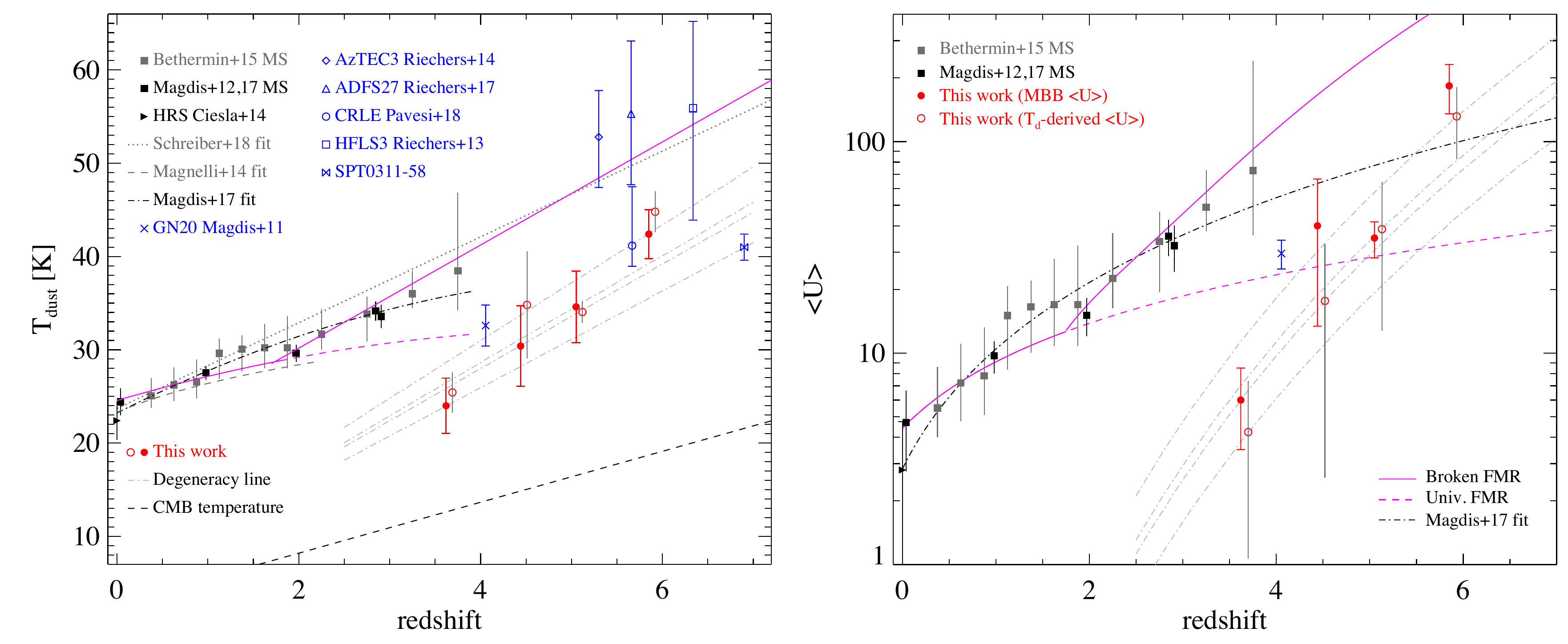}
\caption{%
{Optically thin} dust temperature ${T_{\rm dust}}$ and intensity of the radiation field $\left< U \right>$ versus redshift for galaxies in the literature and the DSFGs studied in this work. 
{\it Left}: Data in this work are marked as red circles, where the filled circles show the MBB results and the open ones show results derived by $\left< U \right>$ scaled to DL07 models. Dust temperatures of MS samples (gray and black filled squares) are converted from $\left< U\right>$ in \cite{Bethermin2015} and \cite{Magdis2012SED,Magdis2017} by  $\left< U \right>= (T_{dust}/18.9)^{6.04}$. The solid and dashed magenta lines represent the evolutionary trends expected for a broken and universal FMR, respectively. 
{\it Right}: Analog to Fig. 5 in \cite{Magdis2017}, overlaid by this work (red circles) and GN20 (blue cross). The filled and open circles correspond to $\left< U\right>$ from MBB SED fitting and $\left< U\right>$ derived from dust temperature according to $\left< U \right>= (T_{\rm dust}/18.9)^{6.04}$, respectively. 
}
\label{Tdust}
\end{figure*}

\begin{figure}
\centering
\includegraphics[width=0.51\textwidth,trim={2.5cm 0.5cm 1.cm 0.cm}, clip]{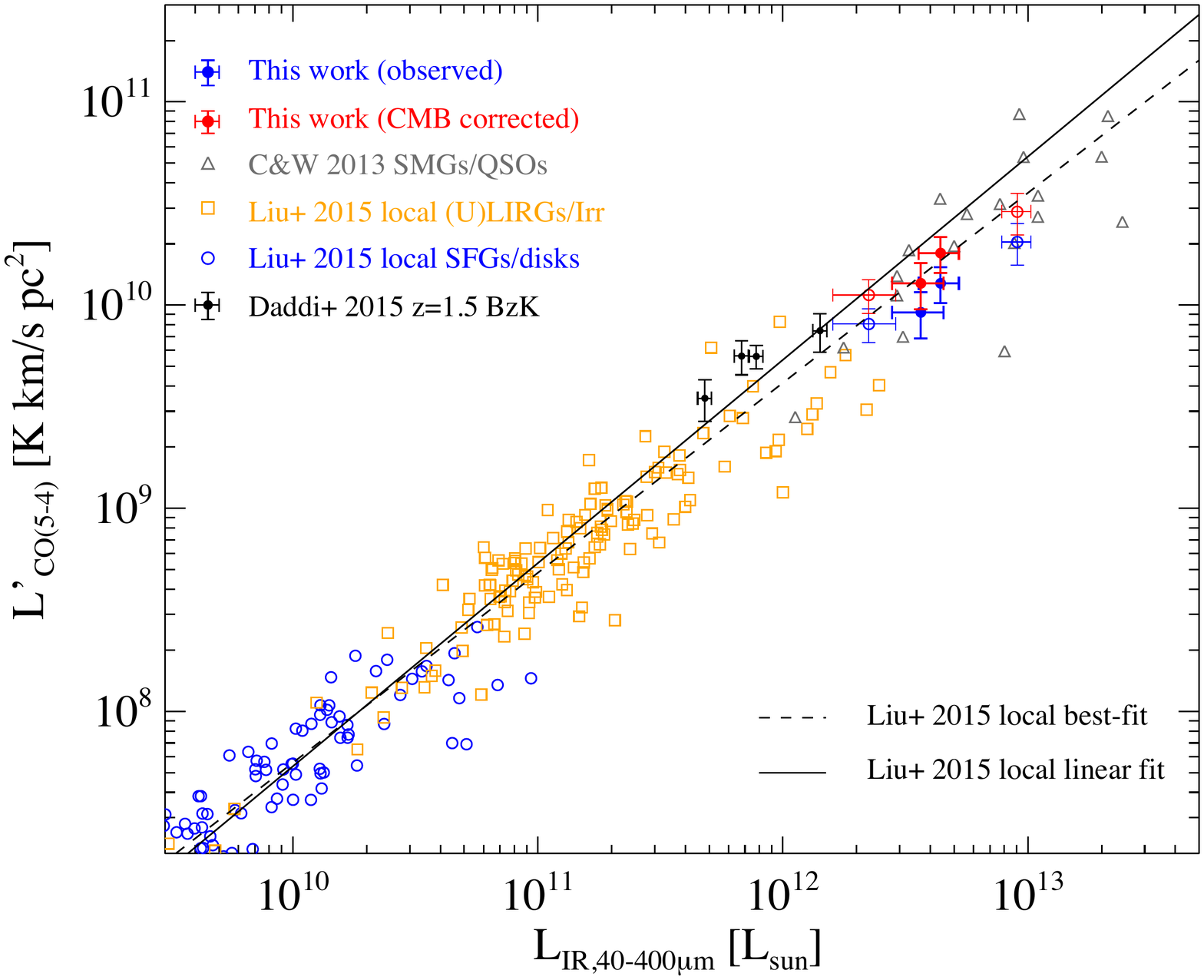}
\caption{%
$L'_{\rm CO}-L_{\rm IR}$ diagram with literature data \citep{Carilli2013,Daddi2015,Liudz2015} and data in this work. Blue filled data points with error bars show observed values in this work, while the red ones show intrinsic values that are corrected for the CMB effect, assuming LTE conditions. {Open circles with error bars are used for the $z=3.623$ and $z=5.847$ galaxies to emphasize that   their $L'_{\rm CO(5-4)}$ were extrapolated as $1.2\times L'_{\rm CO(4-3)}$ and $0.8\times L'_{\rm CO(6-5)}$,   respectively.
Note that a moderate line flux boosting (20\%, see also \citealt{Coogan2018} Fig 5) in this work has been accounted for in the error bar.}
}
\label{CO_IR}
\end{figure}

\subsection{Cold galaxies in the distant Universe}

In general, the impact of CMB on galaxy SEDs can be expressed as a function of the contrast between dust and CMB temperatures, with the highest impact when these temperatures are close. We recall that what counts is the apparent temperature, as encoded in the SED shape and rest-frame peak rather than the intrinsic temperature (which could differ in the case of high optical depths). It is therefore relevant to evaluate how the $T_{\rm dust}$ for our galaxies relates to the cosmic evolution of temperatures in galaxies and the CMB.
In Fig.~\ref{Tdust}-$left$, we compare dust temperature ${T_{\rm dust}}$ for our sample to various objects taken from the literature \citep{Schreiber2018Tdust,Bethermin2015,Magdis2011Dropout24,Riechers2013Nature,Riechers2014,Riechers2017,Pavesi2018,Ciesla2014}, and to the evolving CMB temperature, versus redshift.
We also fitted the available photometry for the $z=6.9$ SPT0311-58 galaxy \citep{Strandet2017,Marrone2017Nature}, and obtained  ${T_{\rm dust}=41\pm2 }$K using the same MBB models including the CMB effect.
{In Fig.~\ref{Tdust}-$right$, we present the comparison in terms of the intensity of the radiation field $\left<U\right>$: all values are consistently and homogeneously measured using the DL07 \citep{Draine2007SED} models, eliminating any systematics in the $T_{\rm dust}$ --$z$ diagram coming from different assumptions of $\beta$ and fitting techniques of the MBB models.}
%
%
In terms of both ${T_{\rm dust}}$ and $\left<U\right>$, this sample has significant cold dust content with respect to both $z<4$ main-sequence galaxies and $z>5$ SMGs.

The $z=5.85$ galaxy has the highest ${T_{\rm dust}}$ and $\left<U\right>$ in this sample, though, its ${T_{\rm dust}}$ is still lower than the extrapolation of the $T_{\rm dust}$--$z$ correlation for MS galaxies, and its $\left<U\right>$ is lower than the extrapolation from a broken fundamental metallicity relation (FMR) as in \cite{Bethermin2015}. 
The $z=5.05$ source shows a dust temperature similar to the $z=4$ starburst GN20 \citep{Daddi2009GN20}. The $z=4.44$ and $z=3.62$ ones have the coldest $T_{\rm dust}$ in this sample, which are just moderately above the CMB temperature by factors of 1.5--2.0.
{Not counting this study, among previously known galaxies, only GN20 is robustly colder than MS galaxies while CRLE is  lower than the MS at only $\sim1.2\sigma$ significance. 
The cold dust in SPT0311-58 at $z=6.9$  had not been reported previously. A significant population of cold galaxies is now clearly detected.}

Please note that increase of ${T_{\rm dust}}$ with redshift within our sample of four targets is simply a selection effect: we had originally selected our targets to be at $z_{\rm phot}>6$ based on photometric redshifts derived using an HFLS3 template. As a result, the lower the redshift with respect to $z=6$, the intrinsically colder the temperature necessarily is {(and larger the implied dust masses)}.
We  show the $T_{\rm dust}$--$z$ degeneracy trends for each galaxy in our sample in  {Fig.~\ref{Tdust}~(and~\ref{Tan}): any redshift trend that our four targets might define is spurious as linked to this degeneracy and affected by selection}. If we had used a  cold dust template like GN20 we would have obtained $z_{\rm phot}=4.0$--5.5 for our targets, which appear to be closer to the observations.

\subsection{CMB impacts observables}
 
We review and summarize here how CMB impacts various observables for this sample, as a necessary step before proceeding to infer their global properties for understanding their nature, and also as a reference for designing and conducting future observations of high-z dusty galaxies.
 
 {\bf Continuum:} 
The CMB acts as a non-negligible background for cold systems against which the line and continuum emission are measured \citep{Combes1999,Papadopoulos2000,daCunha2013}.
According to Eq.~18 in \cite{daCunha2013}, given the observed dust temperatures in our galaxies, the ratios of intrinsic/observed continuum are in the range of 1.32--1.45 at 3~mm and 1.06--1.09 at 1~mm. These ratios are monotonically increasing towards lower redshifts for our sample, an effect due to the fact that lower redshift objects are colder and closer in relative terms to the CMB temperature at each redshift. 
The induced underestimate on the intrinsic dust continuum is more severe at 3~mm, and this has to be carefully accounted for when designing and conducting observations for high redshift galaxies.
The IR luminosities of the galaxies are also somewhat affected by the CMB, with  a small underestimate by a factor of $\approx 1.05$.
 
 {\bf CO emission:} 
A higher CMB temperature enhances the population of the high-J CO levels, and also corresponds to a higher background against which the lines must be detected \citep{Combes1999,Papadopoulos2000,Obreschkow2009}. 
Ultimately for fundamental thermodynamic reasons: $T_{\rm CMB} <= T_{\rm excitation} <= T_{\rm kinetic}$ for any collisionally excited transition.
Then when the kinetic temperature of a cold gas reservoir get closer to the CMB temperature, its low-J CO lines brightness diminish by contrast to the background \citep{Papadopoulos2000,daCunha2013}.
Given the limited range of transitions observed for our galaxies, and the degeneracy between temperature and density in the analysis of {CO SLEDs (e.g., \citealt{Weiss2005CO,Dannerbauer2009CO,Daddi2015})}, it is not possible to properly estimate the expected effect for our galaxies. Nevertheless, one could obtain some guidance by assuming for example that the kinetic temperature of the gas in the CO transitions we observed is close to the dust temperature in the same galaxies, as expected for local thermodynamic equilibrium (LTE). 
This would imply observed CO fluxes {at the level of 70\% of the intrinsic ones according to Eq. 32 in \cite{daCunha2013}. }
This seems to be reasonably consistent with observations. 
In Fig.~\ref{CO_IR} we present the CO(5-4) luminosities of this sample in the $L'_{\rm CO}-L_{\rm IR}$ diagram.
The average ratio of the observed CO(5-4) luminosity (blue data points) to $L_{\rm IR}$ is $\times 0.7$ lower than the best fit and $\times 0.5$ lower than the linear fit in \cite{Liudz2015}.
Thus, the CMB effect provides a reasonable {explanation} for the seemingly low CO fluxes observed in our sample, albeit we cannot fully demonstrate this lacking a measurement of the gas kinetic temperature. This is nevertheless interesting information for planning follow-up searches for CO emission in high redshift galaxies, as reduction factors of up to $\times 2$ in fluxes, due to the CMB, might not be uncommon to be encountered, particularly for  galaxies with low CO excitation temperatures.

{\bf Size measurement:} 
The CMB might also affect the 3mm continuum sizes of our sources, in the case that $T_{\rm dust}$ gradients are present, similarly to local spiral galaxies in which the outskirts are colder. 
In that case, the outskirt surface brightness would be impacted more strongly by the CMB than the hotter centers, resulting in apparently  smaller half light radii.
{Taking as examples the temperature gradients found in some typical local galaxies  M~31, M~33 and NGC~628}, \cite{Zhang_ZY2016} computed that this effect could be at the level of 10--20\% at $z=$4--5 for 3mm bands, while negligible size bias is expected at $\sim1$mm.
Given that this effect is in any case not very strong {(even for surface brightness it would be at most 40\% for the local examples in \citealt{Zhang_ZY2016})}, and that our sources are very compact and possibly merging driven starbursts where we consider it unlikely that they might have strong $T_{\rm dust}$ gradients, we conclude that the CMB should not impact strongly the size measurements for this sample (the situation could be different for extended cold reservoirs in high-z disks). We can verify this directly for  the galaxy  ID85001674 at $z=3.623$. This objects is the coldest and closest in relative terms of the CMB temperature, and shows a combined size $0''.42\pm0''.04$ from 3mm+1mm+0.8mm observations.  The sizes measured from 1mm/0.8mm data alone are $0''.32\pm0''.07$ and $0''.38\pm0''.04$, both about 1--1.5$\sigma$ lower than the combined size. This is contrary to what expected if CMB was affecting sizes, showing that indeed any CMB-induced size bias should be quite small.

{\bf Line widths and dynamical masses:} 
Together with sizes, velocity widths are used to  estimate dynamical masses of galaxies, as we report in Table~\ref{tab:2} ($M_{\rm dyn,2Re}$), following the relations given in \cite{Daddi2010disk,Coogan2018}. 
The CMB  could also impact  the observed CO line width ${\rm V_{FWHM}}$, in the presence of gas kinetic temperature gradients, as described above for the sizes.
The magnitude of this effect has not been quantified so far, but we could expect that it should be roughly comparable to the one on sizes, i.e., probably small. The CO spectra of the galaxies in this work are still relatively low in SNR, and thus we assume that any CMB effect on linewidths would likely be in the noise for our sources.  In any case, the net effect of the CMB on derivation of dynamical mass from observables could imply they are somewhat underestimated given $M_{\rm dyn,2Re} \propto  V_{\rm FWHM}^2\times {\rm R_e}$.
 
Being much less affected by CMB and with higher intrinsic luminosities, [CII]~$158~\mu$m observations would be particularly valuable for this and other high redshift samples. They would provide robust measurements on both size and dynamics, with minimal effects from the CMB. However, see discussion later in Sect.5.2 for other reasons why [CII]~$158~\mu$m might be affected also in these galaxies.

\section{Discussion}

We have presented evidence for the existence of a population of fairly cold dusty star-forming galaxies in the distant Universe, {with SED shapes seemingly much colder (i.e., peaking at longer rest frame wavelengths)} than what is expected for average MS galaxies at the same redshift. We also showed that such cold galaxies are substantially affected by the CMB, so that observed mm fluxes are suppressed by factors up to 1.5, and CO fluxes might be reduced by factors up to 2, or perhaps more for lower-J transitions. This implies that there are observational biases against such cold galaxies, and their real number density might be higher than what has been recovered so far. Up to the point that they might be the majority among SB galaxies in the distant Universe, if the trend determined by \cite{Bethermin2015} for SBs from $z=0$--2 continues to higher redshifts. This raises obvious questions: why are these galaxies so cold? and particularly, why can SB galaxies become systematically colder than MS galaxies, when at $z=0$ the opposite situation is observed? In order to address these questions, we had some closer look at the overall physical properties that can be derived for these galaxies, given the observations available. In physical terms, the $\left<U\right>$ values reflect the ratio of SFE to metallicity. Hence these questions can be rephrased in equivalent terms: are these galaxies low-efficiency star-formers, even lower efficiency than typical MS galaxies? or particularly metal rich? Should we consider them to be typical disk-like star-formers at their redshifts, or might they be merger-affected typical starbursts?

\begin{figure}
\centering
\includegraphics[width=0.5\textwidth, trim={1.5cm 0.4cm 1.cm 1.cm}, clip]{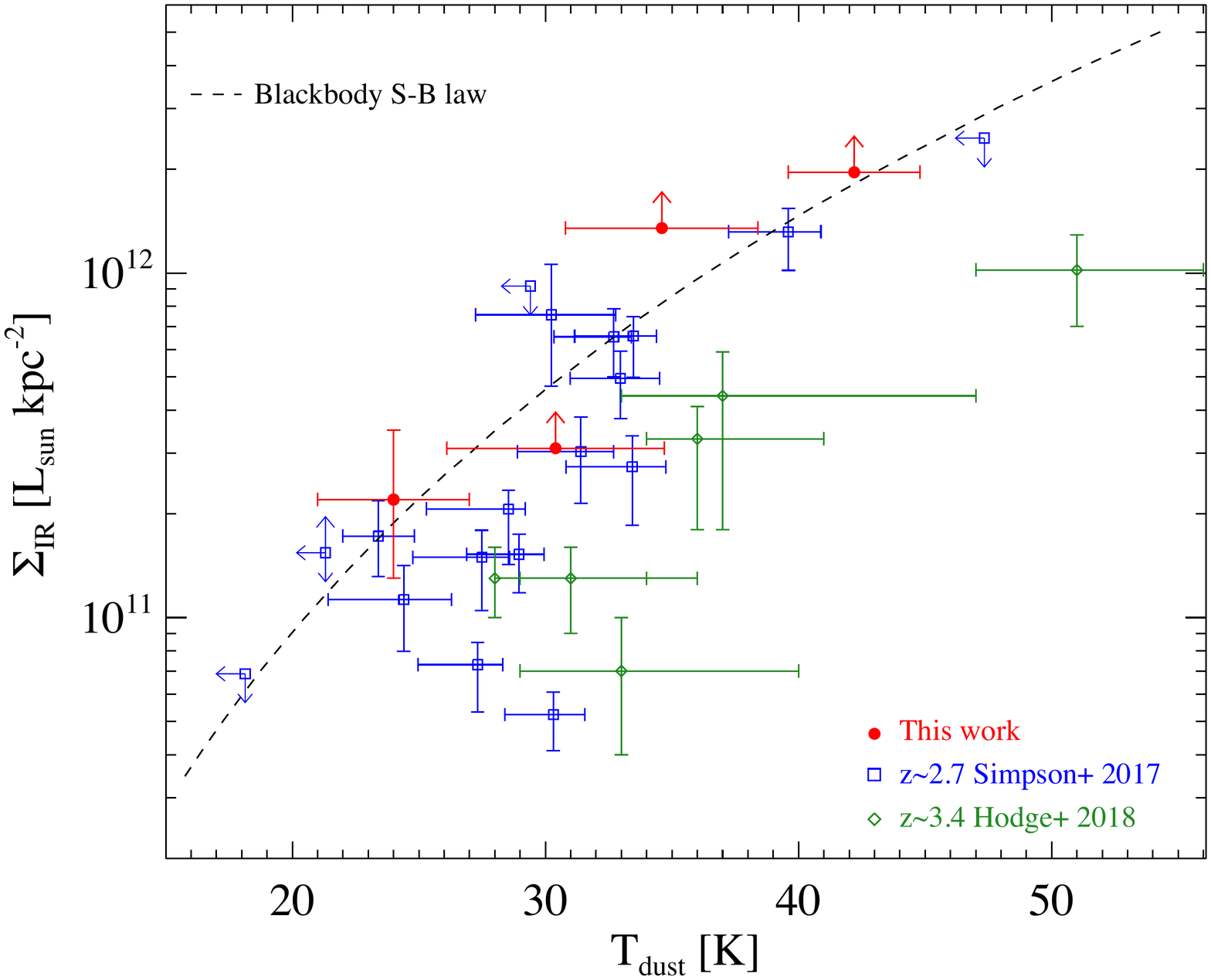}
\caption{%
 IR luminosity surface density versus dust temperature for DSFGs. Blue squares show optically-thin data of SMGs in \cite{Simpson2017}, red diamonds show ALESS sample in \cite{Hodge2018}. Stefan-Boltzmann law is shown as a dashed line, which is only valid for blackbody dust clouds that are optically thick at all FIR wavelengths.
 \label{sigmaSFR}
}
\end{figure}

\begin{figure}
\centering
\includegraphics[width=0.5\textwidth, trim={1.5cm 0.4cm 1.cm 1.cm}, clip]{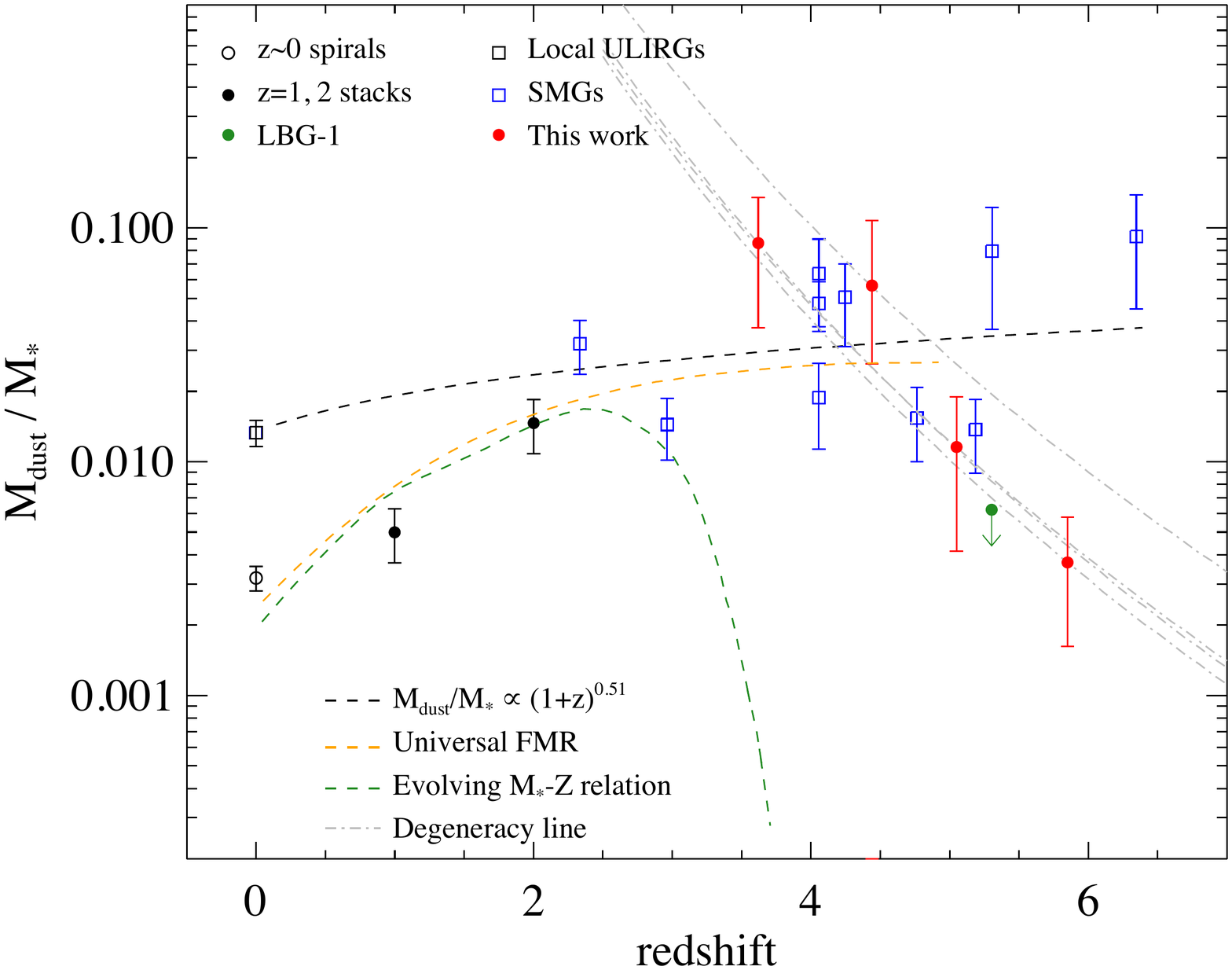}
\caption{%
 Dust to stellar mass ratio versus redshift. Data are taken from \cite{Tan2014} and overlaid by data in this work. 
\label{Tan}
}
\end{figure}

\subsection{Physical properties}

{\bf SFR surface density:} We measure the SFR surface densities by converting the $L_{\rm IR}$ estimates into SFRs \citep{Kennicutt1998KS_law} and dividing by the area $\Sigma_{\rm SFR} = {\rm SFR/(2 \pi {R_{e}}^{2})}$. 
We compared the SFR surface density to  data from the literature, particularly taking samples shown in an analogous figure from \cite{Simpson2017}.
In Fig.~\ref{sigmaSFR}, our sample shows a trend of increasing luminosity surface density with dust temperature, consistent with trends in the lower-z sample of \cite{Simpson2017} and \cite{Hodge2018}. 
Meanwhile our sample shows a higher luminosity surface density than any lower-z sample. Our values are at the limit of the Stefan-Boltzmann law, consistent with a single emitting, homogeneous starburst core. 
{Our sample shows comparable SFR surface density with respect to local (U)LIRGs \citep{Liu_LJ2015}, which are 2--3 orders of magnitude higher $\Sigma_{\rm SFR}$ than normal spirals. The coldest galaxy ($z=3.623$) in this sample appears to be even colder than local IR-detected spirals from  \cite{Liu_LJ2015}, while it has higher $\Sigma_{\rm SFR}$ by 2--3 orders of magnitude.}
The high SFR surface densities suggest these galaxies are rather efficient starbursts than low-efficiency disks. Their extreme compactnesses indeed suggest mergers. In fact, using the evolutionary trend found by \cite{van_der_Wel2014} we would expect that a disk-like galaxy at $z=5$ with $M_{*}\sim5\times10^{10}M_\odot$ to have a FWHM size of $\sim 4.5$~kpc, substantially larger than the sizes we have measured for our objects (Table~\ref{tab:3}).

{\bf Dust and stellar masses:}
MBB dust masses are listed in Table~\ref{tab:3}. We calculated dust mass from DL07 templates, where
we use the intrinsic SED (i.e., corrected for the CMB effect) of each galaxy as recovered from the MBB and took 6 data points (from 80 to 3000~$\mu$m in rest-frame) and fit those data points with DL07. 
The MBB dust masses are consistent within 20\% with the mass scaled to DL07, both result in a massive dust content for this sample.
To constrain their stellar masses, we collected IRAC photometry in literature \citep{Sanders2007} for the $z=5.05$ source, and obtained SPLASH photometry for the others by PSF fitting the SPLASH images on ALMA positions.
We fit stellar SEDs using a 200~Myr template allowing for dust attenuation, following \cite{Liu_DZ2017}. We list stellar masses in Table~\ref{tab:3}. {Despite the IRAC detections well in the optical rest frame we expect uncertainties at least at the level of $\times2$ for our galaxies, given the lack of detections of age dependent features in their optical SEDs. }
In some cases the dust to stellar mass ratios can reach 10\% levels (Fig.~\ref{Tan}), much higher than what is found for typical MS galaxies and consistent with merger-driven systems like bright SMGs (see also \citealt{Tan2014,Rujopakarn2019}).

 \begin{table*}[ht]
{
\caption{Physical quantities}
\label{tab:3}
\centering
\begin{tabular}{cccccccc}
\hline\hline
     ID       &  FWHM size&  $M_*$  &   $M_{\rm dyn, 2Re}$    &  $\Sigma_{\rm SFR}$ &   ${M_{\rm gas,dyn}}$ & ${M_{\rm gas,CI}}$  &  ${M_{\rm gas,FMR}}$ \\
    		 &   $''$ (kpc)   &   [$10^{10}$M$_\odot$]      &      [$10^{10}$M$_\odot$]     &   [${\rm M_\odot yr^{-1} kpc^2}$]  &      [$10^{10}$M$_\odot$]       &       [$10^{10}$M$_\odot$]   &  [$10^{10}$M$_\odot$]    \\
 \hline
85001929  & $<0.33$ (1.90) &     11.6$\pm$5.8 &   $<15.9$    &    $>196$  &  $<8.2$  & -- &	5.7$\pm$1.3 \\
20010161 & $<0.25$ (1.55) &     9.0$\pm$5.2  &    $<12.0$    &    $>135$ &  $<6.8$   & -- &    12.5$\pm$2.8 \\
85000922 & $<0.43$ (2.83)  &     1.8$\pm$1.6 &   $<19.4$   &    $>31$ &  $<16.9$ & -- &   	25.1$\pm$8.6  \\
85001674 & 0.42$\pm$0.04 (3.10$\pm$0.30)  &     3.9$\pm$2.0 &   17.6$\pm$2.6   &    $22_{-9}^{+13}$ & $10.2\pm4.2$  & 9.7$\pm$2.2 & 	48.3$\pm$12.1 \\
\hline
\hline
\end{tabular}\\
}
{Notes: $M_*$: stellar mass; $M_{\rm dyn, 2Re}$: dynamic mass within FWHM size; SFR surface density $\Sigma_{\rm SFR} \equiv {\rm SFR/(2\pi {R_{e}}^{2})}$, where ${\rm R_e}$ is half of the FWHM size. 
} 
\end{table*}

{\bf Starburstiness:} As defined by the ratio of the specific SFR for a given galaxy to the same specific SFR of an average MS galaxy at that redshift. At redshifts $4<z<6$ the MS value is not very securely determined, so this classification is uncertain. However, taking the extrapolation of the trends defined in \cite{Schreiber2015}, {we found that the $z=4.44$ galaxy has a starburstiness SFR/SFR$_{\rm MS}> 4-60$, and appears thus to be a secure starburst galaxy despite its large uncertainty on the stellar mass}.
The remaining galaxies have fairly high masses, and starburstiness of 1.5--2.7, which are all above the MS extrapolation at $z>3.6-6$ but not by large factors. In fact, they could be considered MS galaxies in the definition by \cite{Rodighiero2011}. However, as shown by several works (e.g., \citealt{Elbaz2018_SB_in_MS,Puglisi2019arXiv}), a substantial number of compact, probably merger-driven galaxies are observed even within the MS boundary at high masses, already at intermediate redshift $1<z<3$.

\begin{figure}[ht]
\centering
\includegraphics[width=0.5\textwidth,trim={0.1cm 0.1cm 0.1cm 0.1cm}, clip]{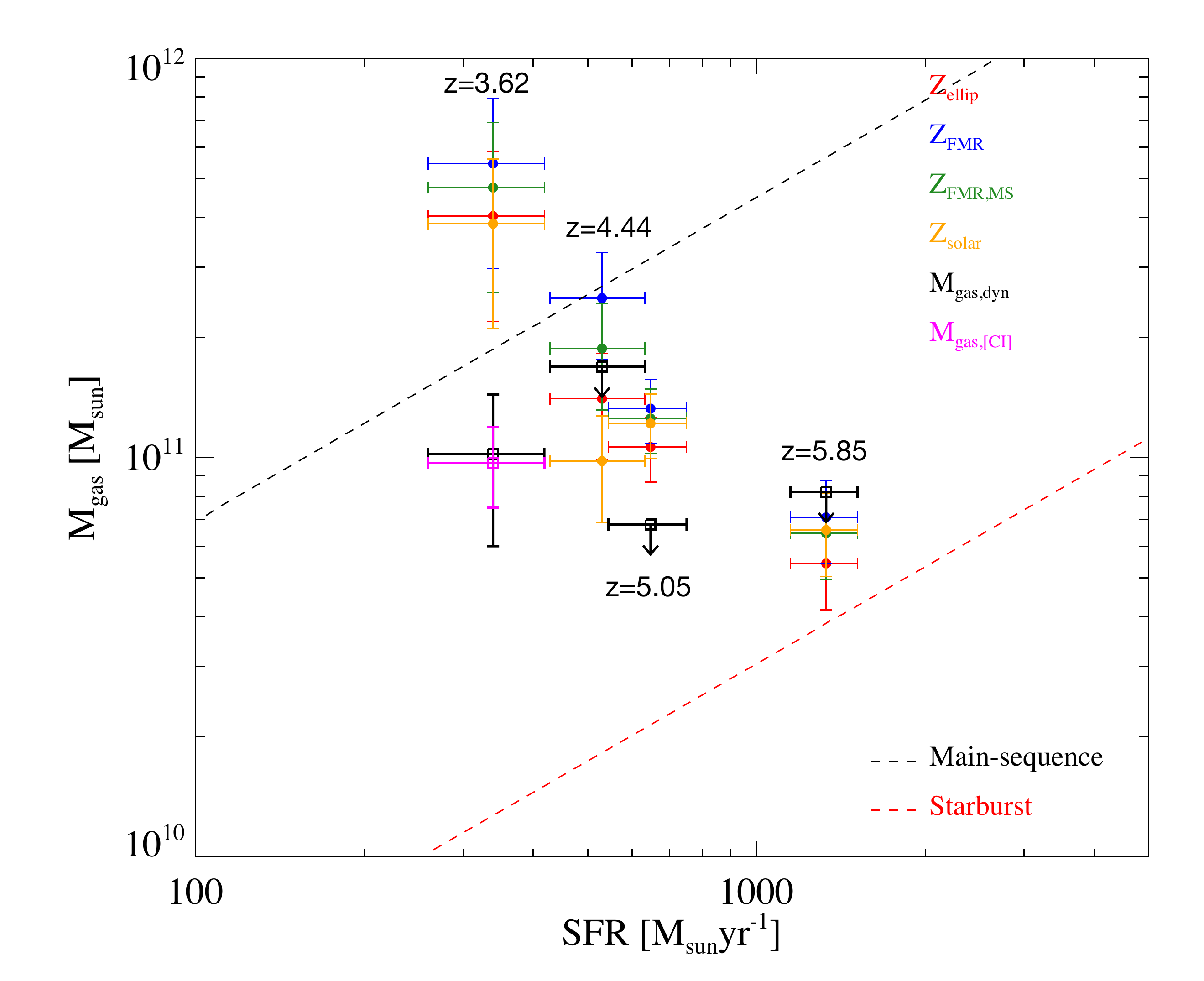}
\caption{%
Gas mass versus SFR for this work. Gas mass is calculated by multiple methods that as labeled (see text for details). 
\label{KS}
}
\end{figure}

{\bf Gas masses:} 
Inferring gas masses would be the most direct way to establish whether these objects are high or low efficiency star-formers. 
We constrain {the total gas mass} (molecular plus any atomic gas) of our sample using 3 different methods: (1) based on the dynamical mass (3 of which are only upper limits), after subtracting stellar mass using the equation 3 in \cite{Daddi2010disk}, assuming $20\%$ of dark matter; (2) based on dust mass scaled to DL07, using FMR, FMR MS, elliptical and solar metallicity based G/D conversion factors, respectively; (3) for the $z=3.62$ galaxy with a [CI](1-0) detection, {given the tight correlations between [CI] and low-J CO in local and high-z galaxies \citep{Valentino2018CI,Jiao2017ci,Jiao2019ci}}, we estimated its gas mass using the relation in \cite{Valentino2018CI} {assuming an excitation temperature of $T_{ex}=$30~K}, and show the resulting ${M_{\rm gas,[CI]}}$ in Table~\ref{tab:3}. 
We cannot estimate gas masses from CO as the lowest-J transition we detect is CO(4-3), whose conversion to CO(1-0) is plagued by too large and uncertain excitation corrections.
In Fig.~\ref{KS}, we compared the 3 different derivations of gas masses in the Schmidt-Kennicutt plane. Comparing to the relations for SBs and MS galaxies taken from \cite{Sargent2014}, one can see that most values tend to be towards the high SFEs typical of SBs, with quite some large range and large uncertainties as exemplified by the scatter in the derived gas masses from different methods and assumptions. 
However it is interesting to consider the $z=3.62$ source which is the coldest in absolute terms and closest to the CMB temperature, and has also alternative derivations of gas mass from the [CI] luminosity and the dynamical mass. 
These two latter estimates agree well between them but are lower than the estimate based on $M_{\rm dust}$ by factors of 4--5, {suggesting that $M_{\rm dust}$ are probably overestimated.}

%

\subsection{Optically thick dust emission in the FIR?}

The behaviour displayed by the galaxy ID85001674, as discussed in the previous subsection, suggests an alternative explanation for the nature of these cold galaxies, which could solve a number of problems with them, particularly the coldest ones. In fact, the dust to stellar mass ratios of the coldest galaxies are very large, reaching 10\% which seems very high {(Fig.~\ref{Tan}~and~\ref{KS})}. Given that the mass in dust can be taken as a proxy of the mass in metals, this would imply that the metallicity in these objects could be truly extreme. Similarly, also the gas to stellar mass fraction would be seemingly $\times 10$ or higher, as is only expected for primordial galaxies, which would be in contradiction with the high metallicities {(Fig.~\ref{KS}}. All these values are derived from $M_{\rm dust}$ estimates, and in all cases assuming models with optically thin dust emission. 
These measurements are not supported by dynamical estimate of the gas masses, nor by derivation from [CI](1-0), a transition known to be in most cases optically thin given the low fractional abundance of  carbon in the neutral form \citep{Valentino2018CI}. It could well be that the dust continuum emission is instead optically thick in these systems, even at the peak of the emission and possibly somewhat beyond, similar to the optically thick SB nuclei found in Arp 220 \citep{Papadopoulos2010Arp220,Wilson2014Arp220,Scoville2017Arp220}.
This is supported by the extreme compactness of our sources, and their very high SFR surface densities. A simple calculation for ID85001674 assuming a spherical geometry with constant density of dust $\rho$ within the measured radius $R_e$, using the measured dust mass from the thin models, and using $\kappa$ from the \cite{Jones2013} opacity models would imply an optical depth ($\tau=\kappa \rho R_e$) at 100$~\mu$m rest (the observed SED peak) towards the center of about 1. 
Of course, there are many implicit assumptions and unknowns in this calculation\footnote{see also the similar calculation for SMG attenuation in SMG by \cite{Simpson2017}, suggesting hundreds of magnitudes of attenuation of the optical rest frame light}, but certainly the case for optically thick dust until the observed SED peak is a plausible one for this galaxies (and for the others, which are likely even more compact).

If the dust emission is optically thick around 100$\mu$m (and possibly beyond), { the suppression of the Wiens emission would make the observed SEDs seem cold}, while the intrinsic temperatures would be much higher (e.g., \citealt{Scoville1976,Condon1991IRAS,Conley2011_lensedSMG,Scoville2012_lecture}). Conversely, together with the underestimate in $T_{\rm dust}$ we would be overestimating $M_{\rm dust}$ by some factor which would depend ultimately on the optical depth (i.e., from the ratio between the intrinsic $T_{\rm dust}$ and the real $T_{\rm dust}$). In all cases $L_{\rm IR}$ (hence SFR) derivations would be unaffected. 

{The standard treatment of optically thick emission assumes a constant $T_{\rm dust}$ at all wavelengths (e.g., \citealt{Casey2012}) and is unlikely to be realistic, given that radial gradients of $T_{\rm dust}$ are expected (e.g., \citealt{Scoville2012_lecture}). Explicitly accounting for such effects is difficult and we hope to present calculations of more physical thick models in future works. Nevertheless we have fitted our targets using the more standard approach of optically thick sources with constant temperature, and accounting for the effects of the CMB. Results are reported in Table~\ref{tab:2} along with the other estimates. The empirical evidence discussed above for ID85001674 requires that we might be overestimating dust masses by factors of up to 4--5 for our sample. These thick models indeed already account for a $\times3$ reduction of $M_{\rm dust}$ for  ID85001674 and $\approx\times2$ for the other galaxies. True corrections might be even larger.}

{If these sources are really optically thick, than the ultimate reason why CMB is so strongly affecting these galaxies is not because their physical dust temperatures are cold but because they {\em seem} cold. Similarly, this would also explain why these galaxies  {\em seem} to be colder than MS galaxies, when they actually might not be in reality. $T_{\rm dust}$ values in Table~\ref{tab:2} derived for the thick fits are already comparable or larger than MS galaxies at the same redshifts. In this case their CO kinetic temperatures would  also be  higher, rendering CMB effects on their line fluxes negligible: the reduction of CO fluxes might be instead due to high optical depths (see also \citealt{Papadopoulos2010Arp220}). }

A strong reduction of $M_{\rm dust}$ in SBs would, finally, also solve the problematic observations that their dust to stellar mass ratios are $\sim5\times$ larger than those of MS galaxies \citep{Bethermin2015}, while dynamical estimates suggest the gas to stellar mass ratios are near unity \citep{Silverman2015,Silverman2018b,Silverman2018a}. The case should be thus seriously considered that many of the merging driven SBs are actually optically thick in the FIR, a case previously proposed for local ULIRGs by \cite{Papadopoulos2010Arp220}.

More observations would be needed to confirm the optically thick case. One interesting path would be comparing to [CI] derived excitation temperatures, given that [CI] lines should remain thin (Cortzen et al 2019 submitted). Another possibility would be to use [CII]$\lambda158\mu$m as a direct tracer of the gas mass \citep{Zanella2018Cii} and from there more realistic $M_{\rm dust}$ estimates via assumptions of G/D ratios: {while [CII] could be in principle heavily affected by optical depth effects, it seems to provide fair gas mass estimates in local ULIRGs (perhaps by lucky coincidence)}. More observations are also needed to establish whether these cold {(i.e., with long peak rest wavelengths)} SEDs are prevalent in the distant Universe, and if the optically thick case could provide a general explanation for the odd inversion that seems to take place at high redshifts, with SB galaxies becoming colder than MS galaxies. 

\section{Summary and conclusions}

We have used ALMA to obtain spectroscopic redshifts and investigate the properties of 4 galaxies detected in the FIR in the {\it super-deblended} catalog by \cite{Jin2018cosmos}, and selected to be at $z>6$ based on photometric redshift derived using a well studied $z=6.3$ galaxy, HFLS3. Our findings can be summarized as follows:

$\bullet$ we securely confirm the redshifts to be at $3.6<z<5.85$ based on multiple CO/CI transitions, with the exception of one object that has only one secure line detected, for which we derive a less secure $z=5.05$.

$\bullet$ our sample contains the most distant spectroscopically confirmed IR-detected galaxy in COSMOS, ID85001929 at $z=5.85$.

$\bullet$ when not accounting for CMB effects the galaxies display highly unusual steep slopes in the Rayleigh-Jeans regimes. The slopes are back to normal when accounting for the CMB. We think this could be the first direct evidence of the impact of the CMB on galaxy observables at high redshifts.

$\bullet$ our galaxies are anomalously cold (i.e., peak at long rest wavelengths), colder than what typical Main Sequence galaxies are expected to be at these redshifts.

$\bullet$ while in general cold dust emission is observed in galaxies with low star formation efficiency, which is typically found in quiescent star forming disks, our galaxies appear instead to be  efficient star formers with compact sizes and  high SFR surface densities, more typical of galaxy mergers.

$\bullet$ we investigate possible reasons for the coldness of the observed FIR spectral energy distributions. An interesting possibility is that the galaxies have optically thick dust emission up until the peak of their emission at about 100~$\mu$m, and possibly beyond. We propose a scheme of high dust optical depths 'hiding' a warm compact dust  mass associated with star-forming, merger-driven 'cores', rather than fitting the associated IR SED with large cold dust reservoirs.

$\bullet$ a large fraction of high (and even low) redshift SBs being optically thick in the FIR would provide an appealing solution into a number of odd observations for this population, including claims of extreme gas fractions and their seemingly redshift-independent $T_{\rm dust}$ evolution. Future work is required to demonstrate whether this is actually the case.

\begin{acknowledgements}
We are grateful to Simone Bianchi and Zhi-Yu Zhang for discussions and to the referee for constructive criticism.
This paper makes use of the following ALMA data: ADS/JAO.ALMA\#2017.1.00373.S, \#2016.1.00463.S and \#2016.1.00279.S. ALMA  is a partnership of ESO (representing its member states), NSF (USA) and NINS (Japan), together with NRC (Canada), MOST and ASIAA (Taiwan), and KASI (Republic of Korea), in cooperation with the Republic of Chile. The Joint ALMA Observatory is operated by 
ESO, AUI/NRAO and NAOJ.
SJ and QG acknowledge supports from the National Key Research and Development Program of China (No. 2017YFA0402703) and the National Natural Science Foundation of China (Key Project No. 11733002). 
SJ acknowledges financial support from the Spanish Ministry of Science, Innovation and Universities (MICIU) under grant AYA2017-84061-P, co-financed by FEDER (European Regional Development Funds).
GEM acknowledges support from  the Villum Fonden research grant 13160 ``Gas to stars, stars to dust: tracing star formation across cosmic time'', the Cosmic Dawn Center of Excellence funded by the Danish National Research Foundation and  the ERC Consolidator Grant funding scheme (project ConTExt, grant number No. 648179)
DL and ES acknowledge funding from the European Research
Council (ERC) under the European Union's Horizon 2020 research and innovation programme (grant agreement No. 694343).
YG is partially supported by National Key Basic Research and Development Program of China (Grant No. 2017YFA0402704), NSFC Grant No. 11861131007 and 11420101002, and Chinese Academy of Sciences Key Research Program of Frontier Sciences (Grant No. QYZDJ-SSW-SLH008).
This paper makes use of the GILDAS software developed by IRAM and available at http://www.iram.fr/IRAMFR/GILDAS.
\end{acknowledgements}

\bibstyle{apj}
\bibliography{biblio}

\end{document}